\documentstyle[12pt]{article}

\catcode`\@=11

\ifcase\@ptsize
 \font\tenmsx=msxm10
 \font\sevenmsx=msxm7
 \font\fivemsx=msxm5
 \font\tenmsy=msym10
 \font\sevenmsy=msym7
 \font\fivemsy=msym5
\or
 \font\tenmsx=msxm10 scaled \magstephalf
 \font\sevenmsx=msxm8
 \font\fivemsx=msxm6
 \font\tenmsy=msym10 scaled \magstephalf
 \font\sevenmsy=msym8
 \font\fivemsy=msym6
\or
 \font\tenmsx=msxm10 scaled \magstep1
 \font\sevenmsx=msxm8
 \font\fivemsx=msxm6
 \font\tenmsy=msym10 scaled \magstep1
 \font\sevenmsy=msym8
 \font\fivemsy=msym6
\fi

\newfam\msxfam
\newfam\msyfam
\textfont\msxfam=\tenmsx  \scriptfont\msxfam=\sevenmsx
  \scriptscriptfont\msxfam=\fivemsx
\textfont\msyfam=\tenmsy  \scriptfont\msyfam=\sevenmsy
  \scriptscriptfont\msyfam=\fivemsy

\def\hexnumber@#1{\ifnum#1<10 \number#1\else
 \ifnum#1=10 A\else\ifnum#1=11 B\else\ifnum#1=12 C\else
 \ifnum#1=13 D\else\ifnum#1=14 E\else\ifnum#1=15 F\fi\fi\fi\fi\fi\fi\fi}

\def\msx@{\hexnumber@\msxfam}
\def\msy@{\hexnumber@\msyfam}
\mathchardef\ltimes="2\msy@6E
\mathchardef\hbar="0\msy@7E
\def\Bbb{\ifmmode\let\next\Bbb@\else
 \def\next{\errmessage{Use \string\Bbb\space only in math mode}}\fi\next}
\def\Bbb@#1{{\Bbb@@{#1}}}
\def\Bbb@@#1{\fam\msyfam#1}

\catcode`\@=12

\newfam\euffam
\font\sixeuf=eufm6
\font\eighteuf=eufm8
\font\twelveeuf=eufm10 scaled\magstep1
	\textfont\euffam=\twelveeuf
	\scriptfont\euffam=\eighteuf
  	\scriptscriptfont\euffam=\sixeuf
\def\euf{\fam\euffam\twelveeuf}

\def\ie{\hbox{\it i.e.}}	
\def\eg{\hbox{\it e.g.}}

\def\Tr{\mathop{\rm Tr}}

\def\dslash{\hbox{$\partial$\kern-1.2ex\raise.2ex\hbox{$/$}}}
\def\Dslash{\hbox{$D$\kern-1.5ex\raise.2ex\hbox{$/$}}}

\def\half{{1\over 2}}
\def\beq{\begin{equation}}
\def\eeq{\end{equation}}
\def\bea{\begin{eqnarray}}
\def\eea{\end{eqnarray}}
\relax

\def\hybrid{\topmargin 0pt	\oddsidemargin 0pt
	\headheight 0pt	\headsep 0pt
	\textheight 9in		% US paper
	\textwidth 6.25in	% A4 paper
	\marginparwidth .875in
	\parskip 5pt plus 1pt	\jot = 1.5ex}

\hybrid

\catcode`\@=11

\def\marginnote#1{}
\newcount\hour
\newcount\minute
\newtoks\amorpm
\hour=\time\divide\hour by60
\minute=\time{\multiply\hour by60 \global\advance\minute by-\hour}
\edef\standardtime{{\ifnum\hour<12 \global\amorpm={am}%
	\else\global\amorpm={pm}\advance\hour by-12 \fi
	\ifnum\hour=0 \hour=12 \fi
	\number\hour:\ifnum\minute<10 0\fi\number\minute\the\amorpm}}
\edef\militarytime{\number\hour:\ifnum\minute<10 0\fi\number\minute}
\def\draftlabel#1{{\@bsphack\if@filesw {\let\thepage\relax
   \xdef\@gtempa{\write\@auxout{\string
      \newlabel{#1}{{\@currentlabel}{\thepage}}}}}\@gtempa
   \if@nobreak \ifvmode\nobreak\fi\fi\fi\@esphack}
	\gdef\@eqnlabel{#1}}
\def\@eqnlabel{}
\def\@vacuum{}
\def\draftmarginnote#1{\marginpar{\raggedright\scriptsize\tt#1}}

\def\draft{\oddsidemargin -.5truein
	\def\@oddfoot{\sl preliminary draft \hfil
	\rm\thepage\hfil\sl\today\quad\militarytime}
	\let\@evenfoot\@oddfoot	\overfullrule 3pt
	\let\label=\draftlabel
	\let\marginnote=\draftmarginnote
   \def\@eqnnum{(\theequation)\rlap{\kern\marginparsep\tt\@eqnlabel}%
\global\let\@eqnlabel\@vacuum}  }

\def\numberbysection{\@addtoreset{equation}{section}
	\def\theequation{\thesection.\arabic{equation}}}

\def\underline#1{\relax\ifmmode\@@underline#1\else
	$\@@underline{\hbox{#1}}$\relax\fi}

\def\titlepage{\@restonecolfalse\if@twocolumn\@restonecoltrue\onecolumn
     \else \newpage \fi \thispagestyle{empty}\c@page\z@
	\def\thefootnote{\fnsymbol{footnote}} }

\def\endtitlepage{\if@restonecol\twocolumn \else \newpage \fi
	\def\thefootnote{\arabic{footnote}}
	\setcounter{footnote}{0}}  %\c@footnote\z@ }

\catcode`@=12
\relax

\newskip\humongous \humongous=0pt plus 1000pt minus 1000pt

\newif\ifdtup

\relax

\def\wt{\widetilde}
\def\tgu{\widetilde{g}(u)}
\def\K{K(h_\psi,h_\psi)}
\def\x{(x)}
\def\u{(u)}
\def\xu{(u)}
\def\dbar{\bar{\partial}}
\def\gt{\widetilde{g}}
\def\qb{\bar{q}}
\def\zb{\bar{z}}
\def\H{{\cal H}}
\def\ad{\mathop{\rm ad}}
\def\rank{\mathop{\rm rank}}
\def\cox{ h_{\g}}
\def\g{{\euf g}}
\def\t{{\euf t}}
\def\m{{\euf m}}
\def\ip#1#2{\left(#1,#2\right)}
\def\IP#1#2{\left\langle#1,#2\right\rangle}
\def\d{\partial}
\def\db{\bar{\partial}}
\def\ham{P_0}
\def\mom{ P_1}
\def\l{L_0}
\def\lb{\bar{L}_0}
\def\lgt{\widetilde{LG}}
\def\waff{W_{\rm aff}}	%affine Weyl group
\def\ewaff{w_{\rm aff}}	%element of affine Weyl group
\def\cmu{\check{\mu}}
\def\ct{\check{T}}
\def\phit{\widetilde{\varphi}}
\def\gamt{\widetilde{\gamma}}
\def\tw{\theta_w}
\def\vth{\vartheta_{11}}
\def\subchi#1{\lower1ex\hbox{$\scriptstyle#1$}}
\def\Fzero{{\cal F}_0}
\def\DET{\mathop{\rm DET}}
\def\GC{G_{\Bbb{C}}}

\def\th{\widehat{\tau}}
\def\ghor{g_{h}}
\def\gamhor{\gamma_{h}}
\def\half{{1\over 2}}
\def\Omtt{\widetilde{\widetilde{\Omega}}}
\def\TT#1{\widetilde{\widetilde{#1}}}
\def\L{{\cal L}}

\numberbysection
\begin{document}
\begin{titlepage}
\noindent
September 1991 \hfill UCB-PTH 91/37\newline
\strut\hfill LBL--31093\newline
\strut\hfill LPTHE 91--43
\par\vskip .3in
\begin{center}
\large\bf The Supersymmetric $\sigma$-Model and the Geometry\\
	of the Weyl-K\v{a}c Character Formula\footnote{This work was
supported in part by the Director, Office of Energy Research, Office of
High Energy and Nuclear Physics, Division of High Energy Physics of the
U.S. Department of Energy under contract DE-AC03-76SF00098, in part by
the Division
of Applied Mathematics of the U.S. Department of Energy under contract
DE-FG02-88ER25066, in part by the National Science Foundation under
grant PHY-90-21139, and in part by the Centre National de la Recherche
Scientifique.}\normalsize\rm \\[.3in]
{\normalsize\bf Orlando Alvarez}\\
\small\it Department of Physics\\
University of California at Berkeley\\
and\\
Theoretical Physics Group\\
Lawrence Berkeley Laboratory\\
Berkeley, CA 94720, USA\\[.1in]
{\normalsize\bf I.M. Singer}\\
Department of Mathematics\\
Massachusetts Institute of Technology\\
Cambridge, MA 02139, USA\\[.1in]
{\normalsize\bf Paul Windey}\\
LPTHE\footnote{Laboratoire associ\'e No. 280 au CNRS.}\\
Universit\'e Pierre et Marie Curie\\
Tour 16, 1er \'etage\\
4 Place Jussieu\\
F--75252 Paris CEDEX 05, FRANCE\/ \normalsize\rm \\
\end{center}
\begin{abstract}
Field theoretic and geometric ideas are used to construct a chiral
supersymmetric field theory whose ground state is a specified irreducible
representation of a centrally extended loop group. The character index of
the associated supercharge (an appropriate Dirac operator on $LG/T$) is the
Weyl-K\v{a}c character formula which we compute explicitly by the steepest
descent approximation.
\end{abstract}
\end{titlepage}

\section{Introduction}

Many results in Lie group theory have both an algebraic and a
geometrical origin.  The Weyl character formula~\cite{weyl.group} is no
exception and among the many proofs of it some are purely algebraic in
nature, others purely geometrical.  This complementarity extends often
to the case of infinite dimensional loop groups with central extension
(for example, \cite{kac.book,pressley-segal.loopgroups,f-l-m.book}).
In view of the current importance of loop groups and of the related
affine Lie algebras in theoretical physics it is useful to have, as
much as possible, independent proofs of important results based on
physical methods.  It is well known from past experience in quantum
mechanics that the operator formalism is best suited to obtain
algebraic insight while the path integral is better apt to reveal the
geometrical foundations of a particular theory.  We adopt this
viewpoint to derive the central result of this paper which is the
Weyl-K\v{a}c character formula.  The motivation is to get a better
geometrical understanding of some related aspects of loop groups and
conformal field theory.  To develop the necessary machinery in field
theory it is always useful to have as much as possible a quantum
mechanical analogy.  The corresponding derivation of the classical Weyl
character formula was previously published separately \cite{asw.weyl}
and will be referred to as~I throughout the rest of this paper. The
reader unfamiliar with the material of Section~\ref{review} is
encouraged to consult it for a pedagogical introduction to the methods
used here.

We obtain the Weyl-K\v{a}c character formula by computing the index of
a certain Dirac operator on an infinite dimensional manifold. The index
obtained is  different from previous elliptic genus computations
\cite{akmw.irvine,akmw.cmp,p-s-w,witten.genera,witten.loop} which were
associated with Dirac operators on $LM$, the loop space of a finite
dimensional manifold $M$. The elliptic genus from the algebraic
topology viewpoint is discussed in \cite{landweber.book}. Here we
discuss the Dirac operator on $LG/T$, a homogeneous space naturally
associated to a connected, simply connected,  simple, compact Lie group
$G$ with maximal torus $T$. The space $LG/T$ is not the loop space of
any manifold. However, there is still an $S^1$ action on it which plays
an important role since it is responsible for the affine grading of the
character.

The paper is organized as follows.  In Section~\ref{review} we review
the Borel-Weil construction of the representations of a loop group
$LG$. The representations are obtained as holomorphic sections of line
bundles over the coset space $LG/T$. We also explain the crucial role of
$\widetilde{LG}$, the central extension of $LG$. In particular we
discuss how the central extension may be seen as a $U(1)$ bundle over
$LG$ \cite{pressley-segal.loopgroups}, an interpretation which is
important for our purpose. We also discuss the generic features of the
construction of Atiyah and Bott~\cite{a-b.1,a-b.2} which links group
characters with fixed point formulas and character indices. This
analysis permits us to identify the correct physical theory whose
supersymmetric ground states will realize the representations of the
loop group. We also introduce the partition function which will give
the character formula for loop groups.

The explicit realization of these ideas in the framework of a very
special field theory with chiral supersymmetry as well as the
construction of its lagrangian is developed in
Section~\ref{lagrangian}. There we introduce the coupling of
``matter'', \ie\/ the $T$ gauge couplings which correspond to  the
Borel-Weil bundles mentioned above.  This entire construction requires
a delicate extension of the concept of {\em horizontal supersymmetry\/}
already introduced in~I.

Finally the explicit computation of the Weyl-K\v{a}c character formula is
detailed in Section~\ref{computation}. Most of our notational conventions
are defined in Appendix~A and we will use them freely without further notice.

Bernard~\cite{bernard.wzw} has derived the Weyl-K\v{a}c character formula
by using a mixture of conformal field theory and mathematical results. He
uses mixed Virasoro --- affine Lie algebra Ward identities in the WZW
model~\cite{witten.wzw}, properties of the Macdonald identities derived
without using affine algebras, and a variety of results related to the heat
kernel on $G$. A field theoretic purely algebraic construction of the
Weyl-K\v{a}c character formula was presented by Bouwknegt, McCarthy and
Pilch~\cite{b-m-p.freefield}.
These authors apply the Euler-Poincar\'e-Lefshetz principle to certain free
field Fock spaces which are built with the aid of BRST operators associated
with ``screening currents''. Warner~\cite{warner.kac} has employed
supersymmetric index technology to give a proof of the Weyl-K\v{a}c
character formula along more algebraic lines.

\section{Borel-Weil Theory and Further Preliminaries}
\label{review}

We now  briefly set the stage for our problem
(a complete treatment
in the spirit of this paper was given in I for the ordinary Lie
group case). Firstly we will build the representations of
the loop group following the method of Borel-Weil:
\begin{itemize}
\item to each irreducible representation we associate its infinitesimal
character in the maximal torus of the group (essentially the
highest weight of the representation);
\item this character uniquely defines a line bundle over the complex
manifold formed by the coset space of the group over its maximal torus;
\item the holomorphic sections of the line bundle
 provide an explicit construction of the representation.
\end{itemize}
The group we have to consider is $\widetilde{LG}$, the central
extension by $U(1)$ of the loop group $LG$ (which locally looks like
$LG \times U(1)$). The multiplication of two elements (in local
coordinates) $(g(x),u)$ and $(g'(x),v)$ of $\widetilde{LG}$ is given
by
\beq
(g(x),u)(g'(x),v)=(g(x)g'(x),uv\Phi(g,g'))\;,
\eeq
where $\Phi$ denotes the cocycle associated with the $U(1)$ central
extension of the loop group. In this case the steps outlined above
construct the line bundle ${\cal L}_{(\lambda,k)}$ over
$\widetilde{LG}/(T\times U(1))$ associated with a character $(\lambda,k)$
of $T\times U(1)$. Three important remarks are in order here. Firstly, the
space $\widetilde{LG}/(T\times U(1))$ is isomorphic to $LG/T$. Secondly one
can view the central extension $\widetilde{LG}$ as a special $U(1)$ bundle
over $LG$\footnote{For more details on these questions we refer the reader
to the excellent exposition given in the book of Pressley and Segal
\cite{pressley-segal.loopgroups}.
We remind the reader that the Lie group $G$ is connected, simply connected,
simple and compact.}. Thirdly, the special line bundle $\L$ over $LG/T$
arising from the basic central extension of $LG$ (as discussed in the
Introduction) is isomorphic to $\L_{(0,1)}$. The transposition of these
ideas to a physical context is {\it a priori\/} straightforward: we
construct a quantum mechanical system whose configuration space is the
coset manifold $LG/T$ and couple it to an external gauge field
corresponding to the group $T\times U(1)$ via an ordinary minimal coupling
$A_\mu\dot{x}^\mu$. The wave functions of this system will be sections of
$\L_{(\lambda,k)}$. The subtlety lies in the proper identification of the
$U(1)$ part of this coupling. To elucidate this point it is best to view a
quantum mechanical system over $LG$ as a non-linear two dimensional
$\sigma$-model with group $G$. It can then be shown that the central $U(1)$
coupling corresponds to the addition of the Wess-Zumino term (see below.)
We still have to quotient by $T\times U(1)$. This amounts to choosing an
appropriate connection as we will see in Section~\ref{computation}.

The second main ingredient we need is, {\it mutatis mutandis\/}, the
Atiyah-Bott construction \cite{a-b.1,a-b.2}.  Since the irreducible
representation coincides with the holomorphic sections of
$\L_{(\lambda,k)}$, it is clear that they belong to the kernel of
$\db\otimes  I_{\L_{(\lambda,k)}}$, the $\db$ operator over $LG/T$
twisted by the  line bundle $\L_{(\lambda,k)}$.  The computation of the
index of this operator should in principle give us the dimension of the
representation\footnote{Actually one should prove a vanishing theorem
since the irreducible representation is given by the cohomology group
$H^{(0,0)}(LG/T,{\L_{(\lambda,k)}})$ and the rest of the cohomology
groups are required to vanish.} while to find the character of the
irreducible representation we have to compute the character index. We
can equivalently work with the Dirac operator $\dslash$ provided we
compensate by an
extra twist (see I) to make up for the difference between the Dirac
operator and $\db$.

The analogous statements in our physical setting are familiar: we first
construct the supersymmetric extension of the model (the generator of
supersymmetry is identified with the Dirac operator\footnote{The Dirac
operator we consider is the naive Dirac operator plus Clifford
multiplication by the natural $S^1$ vector field~\cite{witten.argonne}
plus appropriate gauge couplings.} on the configuration space) and then
compute $\Tr(-1)^Fg$  to
obtain the character index
formula~\cite{goodman-witten.index,goodman.index}.
Here   $(-1)^F$ is the fermion parity operator and $g\in T$.
Again this
construction carries over to the loop group case; one simply works with
the Dirac-Ramond operator $\bar{G}_0$ with appropriate gauge couplings.
 The Dirac-Ramond operator is the generator of supersymmetry in our
$\sigma$-model. Notice that the above discussion implies that one
should have only one supersymmetry generator: our construction will be
chiral in an intrinsic way.  The general principles involved in the
computation of the index of the Dirac-Ramond operator are by now
standard from the work on elliptic
genera~\cite{akmw.irvine,akmw.cmp,p-s-w,witten.genera,witten.loop}.
The main task facing us is thus the construction of the lagrangian
germane to the situation we have just analyzed. Let us warn the reader
that the description just given is very sketchy, in particular we will
see below and in the next section that the naive $\sigma$-model
lagrangian is completely unacceptable before we add the Wess-Zumino
term. The introduction of the central extension is forced by reasons of
symmetry. All these details as well as the boundary conditions will be
discussed at length below and in the next section.

Next we note that there is a left action by the
maximal torus of the group (here $T\times U(1)$) on the coset space. The
fixed points of this action are the affine Weyl group $\waff$ defined
in Section~\ref{computation}.  In the loop group case this fixed point
set is infinite; still we expect the index computation to reduce to a
neighborhood of the fixed point set~\cite{pressley-segal.loopgroups}.
The first step in implementing these ideas should be the
construction of a lagrangian which admits the loop group $LG$ as a
group of symmetries. As previously mentioned, let us see why the obvious
attempt at constructing such a lagrangian fails. The simplest choice is
the standard $(1+1)$-dimensional nonlinear $\sigma$-model defined by a
map $g: \Sigma \to G$ where the world sheet $\Sigma$ will be taken to
be a torus. The dynamics of the  model defined by the classical action
\beq
	\label{NLS}
	\int_\Sigma d^2z\; \Tr (g^{-1} \partial_a g) (g^{-1} \partial_a g)
\eeq
may be interpreted as the motion of
a ``particle'' on the loop group $LG$. Unfortunately, (\ref{NLS}) does
not have a large enough symmetry group for our purposes. The classical
action is not invariant under the action by the loop group $LG$. In
fact, the symmetry group of (\ref{NLS})  is $G \times G$ where the group
action is defined by $ g(x,y) \mapsto h_L g(x,y) h_R^{-1}$ for
$(h_L,h_R) \in G\times G$. The hamiltonian defined by (\ref{NLS}) does
not have the $LG$ symmetry we require. However, it is well known
\cite{witten.wzw} that the addition of the Wess-Zumino term to the
lagrangian (\ref{NLS}) extends the symmetry group to $LG\times LG$ (in
Minkowski space).
The Wess-Zumino-Witten (WZW) action~\cite{witten.wzw} reads
\bea
 I_{\rm WZW} &=& { k \over 6 \pi i \K}
	\left( \vphantom{\int_\Sigma \int_B}\right.
	- 6i \int_\Sigma d^2z\; \Tr\left\{
	(g^{-1}  \d_z g)(g^{-1}\d_{\zb}g)
	- \gamma \d_z \gamma
	+ (g^{-1} \d_z g)\gamma\gamma \right\} \nonumber\\
	\label{WZW}
	&+&\left. \vphantom{\int_\Sigma \int_B}
	\left[
	 \int_B \Tr(g^{-1} dg)^3
	+ 6i \int_\Sigma d^2 z\;\Tr( g^{-1} \d_z g)\gamma\gamma \right]
	\right)\;,
\eea
where $B$ is a three manifold such that $\d B=g(\Sigma)$,
$\Tr$ stands for $\displaystyle \Tr_{\rm ad}$,
and $k$ is a
positive integer; see Appendix~A for the notation.
This model is conformally invariant and also formally admits $L\GC
\times L\GC$ as a symmetry group, where $\GC$ is the complexification
of the Lie group $G$.  To be more precise one has the
following formal symmetry of the action: $ g(z,\zb) \mapsto h_L(z)
g(z,\zb) h_R(\zb)^{-1}$ where we think of the left action as being
generated by locally holomorphic maps into $\GC$ and the right action
being generated by locally antiholomorphic maps into $\GC$.  For many
practical purposes we may think of $L\GC$ as analytic maps of an annulus
into the group $\GC$. The relative normalization of the kinetic energy
term of (\ref{WZW}) and the Wess-Zumino term was forced on us by
demanding that we choose a lagrangian which admits an $LG\times LG$
symmetry\footnote{Throughout this article we will follow the
physics convention of referring to this symmetry as $LG \times LG$.
A further discussion of $LG$ versus $L\GC$ will be given shortly.}.

We now explain the phrase ``Wess-Zumino'' term which is liberally used
throughout this article. A Wess-Zumino term is a special case of the
following general set up. Assume $M$ is a connected, simply connected
manifold with a line bundle with connection~$A$. The
lagrangian describing the motion of a particle (on the base $M$) moving
in the presence of the connection $A$ will generically have three types
of terms: kinetic energy terms, potential energy terms and a gauge
coupling term. We are interested in the gauge coupling term. Assume a
path $\gamma$ begins at a point $u_0\in M$ and ends at $u_1$. The gauge
coupling term contribution to the path integral is simply
\beq
	\label{par-trans}
	\exp \int_\gamma A\;,
\eeq
\ie, parallel transport from $u_0$ to $u_1$ along $\gamma$. Note that
this object transforms ``bi-locally'' under a gauge transformation.
When $\gamma$ is a loop, the simply connected nature of $M$ tells us that
$\gamma = \d D$ where $D$ is a disk. In this case we see that
\beq
	\exp \int_\gamma A = \exp \int_D F\;
\eeq
where $F= dA$ is the curvature. Thus for loops, the lagrangian may be
formulated in terms of curvature. Note that $\exp \int_D F$ is
independent of choice of disk because the first Chern class of a line
bundle $\int iF/2\pi$ is integral.  In the case where $M=LG$ the term
$\int_D F$ is called a Wess-Zumino term.

Although we have to consider open paths, we will be able to formulate
the problem in terms of curvature which simplifies calculations. Our
path integral calculation is dominated by the critical points of the
steepest descent approximation. Since the result is given exactly by the
quadratic approximation all we have to do is understand what happens in
a neighborhood of a critical point $u_c$. Pick a family $\{\Gamma(u)\}$
of fiducial paths connecting the origin $u_c$ to a point $u$ in the
neighborhood. If $\gamma$ is a path connecting the initial point $u_0$
to the final point $u_1$ then the gauge coupling term may be
written as
\beq
	\label{WZ}
	\exp \int_\gamma A = \exp \left\{ \int_{D(\gamma)} F
	- \int_{\Gamma(u_0)} A + \int_{\Gamma(u_1)} A \right\}\;,
\eeq
Where $D(\gamma)$ is a disk with boundary given by the loop
$\Gamma(u_0)\circ\gamma\circ\Gamma(u_1)^{-1}$.
In the steepest descent approximation to the path integral, we have to sum
over all paths $\gamma$ in the neighborhood of the critical point $u_c$.
As the path $\gamma$ varies, the only term in (\ref{WZ}) which changes
is the curvature term. The line integrals along $\Gamma(u_0)$ and
$\Gamma(u_1)$ are there to enforce the gauge transformation properties
of parallel transport. Remember that the curvature term is gauge
invariant.
The situation is actually a bit better as we will see later
in this section.

Even though the correct coupling is (\ref{par-trans}) we will abuse the
situation and write it as a Wess-Zumino curvature term. The
justification is two fold. First, we have the discussion of the
previous paragraph. Second, in the case of a loop group, it is very easy
to write the curvature yet the expression for the connection is neither
nice nor illuminating. From now on we will blur the distinction between
a Wess-Zumino term and the correct gauge coupling. We interpret a
Wess-Zumino term as parallel transport when necessary.

The connection to Borel-Weil theory will require a supersymmetric model
and in anticipation we have included a $(0,1/2)$ fermion $\gamma$ in
(\ref{WZW}).  The fermion $\gamma(z,\zb)$ is a {\it left invariant\/}
element of the Lie algebra of $G$ and is the superpartner of $g$ (see
Section~\ref{lagrangian}). Equation~(\ref{WZW}) is a chiral $(0,1)$
supersymmetric extension of the ordinary WZW action\footnote{A
non-chiral $(1,1)$ supersymmetric WZW was first studied
in~\cite{d-k-p-r.susywzw}. The model discussed here is quite
different.}.
Notice that a term of the form $(g^{-1}\d_z
g)\gamma\gamma$ does not appear in (\ref{WZW}) due to a cancellation
between the contribution in the curly braces and the one in the square
brackets. The curly braces expression and the square brackets
expression are each independently supersymmetric. In fact, the term in
curly braces is the generalization of equation (4.15) of I to field
theory. The term in square brackets is the supersymmetric version of
the Wess-Zumino term. It is of the form $A \dot{x} + F \psi\psi$
discussed in I (actually the $A \dot{x}$ term is written as a curvature
term). Later in this section we will see that the curvature is given by
(\ref{Lambda-def}) and thus $(g^{-1}\d_z g)\gamma\gamma$ is of the
$F\psi\psi$ type.

The classical equations of motion are
\bea
	\d_z\left( g^{-1} \d_{\zb} g \right) &=& 0\;,\\
	\d_z \gamma &=& 0\;.
\eea
Action (\ref{WZW}) is invariant under the supersymmetry transformations:
\bea
	g^{-1} \delta_s g &=& \varepsilon\gamma\;,\\
	\delta_s \gamma &=& \varepsilon(g^{-1}\d_{\zb} g
		- \gamma\gamma)\;,
\eea
where $\varepsilon$ is an anticommuting parameter. The associated
supercurrent has conformal weight $(0,3/2)$ and is defined by
\beq
	{\cal S} \propto \Tr\left[ \left(g^{-1}\d_{\zb}g\right) \gamma
		+ \gamma\gamma\gamma \right]\;.
\eeq
The two current algebras associated with $LG\times LG$ are given
by
\bea
	\label{Jbar}
	J_{\zb} &=&{2k\over K(h_\psi,h_\psi)} \left(g^{-1}\d_{\zb}g
	+ \gamma\gamma\right) \;,\\
	J_z &=&{-2k\over K(h_\psi,h_\psi)}\; \left(\d_z g\right) g^{-1}\;.
\eea
Note that $J_{\zb}$ generates the right group action and $J_z$
generates the left group action.
 To each element $X\in L\g$, the Lie algebra of $LG$, we associate
the operator
\bea
\label{currents}
J_X&=&-\oint {dz\over 2\pi i} K(X(z),J(z))\\
&=&\oint {dz\over 2\pi i} X^a(z)J_a(z)\\
&=&2k\oint {dz\over 2\pi i} {K(\partial_zgg^{-1},X)\over
K(h_\psi,h_\psi)}\; .
\eea
The affine algebra is given by the  commutation relations:
\beq
\label{k-m}
\left[ J_X,J_Y\right]=J_{[X,Y]}- 2k \;\oint {dz\over 2\pi i}\;
{K\left(X(z),{dY(z)\over dz}\right) \over K(h_\psi,h_\psi)}\; ,
\eeq
or  in an orthonormal basis where $K(e_a,e_b)=-\delta_{ab}$ and
$[e_a,e_b]=f_{ab}{}^c e_c$ define the  structure constants, the associated
operator product expansion is
\beq
J_a(z)J_b(w)\sim - {\delta_{ab}\over K(h_\psi,h_\psi)}{2k\over (z-w)^2}
+ {f_{ab}{}^c J_c(w)\over (z-w)}\; .
\eeq

We now return to the discussion of $LG$ and $L\GC$. For simplicity we
will temporarily assume that the worldsheet $\Sigma$ is either
Minkowski or Euclidean space and only provide a ``local description''.
The confusion  in whether to write $LG$ or $L\GC$ arises in the Wick
rotation from Minkowski space to Euclidean space. The physical world
sheet is Minkowski space. The analogues of complex coordinates are the
light cone coordinates $x^{\pm} = x \pm t$ where $x$ is the spatial
coordinate and $t$ is the temporal coordinate. In terms of these
coordinates, the symmetry of the WZW model is $g(x,t) \mapsto h_L(x^-)
g(x,t) h_R(x^+)$. We immediately see that $h_L$ and $h_R$ are functions
of a single real variable. If we take the worldsheet to be a Minkowski
cylinder then we will have a legitimate $LG\times LG$ symmetry.  When
we Wick rotate to Euclidean space $x^- \to z$ and $x^+ \to \zb$, so
$h_L(x^-)$ and $h_R(x^+)$ become functions of $z$ and $\zb$
respectively; thus one has to complexify and look at analytic and
anti-analytic maps into  $\GC$. The following observation illustrates
the nature of $h_L(z)$.  Consider the standard mode expansion of the
current operators in conformal field theory
\beq
	J_a(z) = \sum_{n=-\infty}^\infty {J_{a,n} \over z^{n+1}}\;.
\eeq
The hermiticity of the currents in Minkowski space translates into the
operator relations $J_{a,n}^\dagger = J_{a,-n}$ in the conformal field
theory. The operator
\beq
	\exp J_X = \exp \oint {dz \over 2\pi i}\; X^a(z) J_a(z)
\eeq
can be a unitary operator on the Hilbert space
if $X(z)$ is chosen appropriately.
If in the mode expansion $X^a(z) = \sum_{-\infty}^\infty X^a_n/z^n$
we require $X^a_{-n}
= - \overline{X^a_n}$ then $J_X$ is antihermitian and one formally gets a
unitary operator $\exp J_X$ on the Hilbert space. It is in this sense
that one has a map into $LG$, more precisely, a unitary representation
of the centrally extended loop group. Such a $X(z)$, which in general is
not an analytic function, may be formally considered a map of the
annulus into $\g$. Note that on $|z|=1$, $X^a(z)$ is pure imaginary and
thus define via exponentiation a map into $LG$. Often in physics one
concentrates collectively on the basis $\{J_{a,n}\}$
and thus the distinction of whether one is working on $L\g$ or
$L\g_{\Bbb C}$ is blurred.

For our purposes we will need a different interpretation of the Wess-Zumino
term in (\ref{WZW}). Notice that this term is first order in the time
derivatives and thus is of the $A_\mu(x)\dot{x}^\mu$ form previously
mentioned and also described in detail in I.
Equivalently, the centrally extended loop group $\lgt$ may be interpreted
as a $U(1)$ bundle over $LG$, see for
example~\cite{pressley-segal.loopgroups}.  The WZW action describes the
motion of a superparticle on $LG$ in the presence of a $U(1)$ gauge
potential; therefore the quantum mechanical wavefunction for this system is
a section of a line bundle over $LG$ with first Chern class $k$.

In summary, we found a supersymmetric action $I_{\rm WZW}$ for a
superparticle moving on $LG$ which admits $LG\times LG$ as a symmetry
group ---  this is still too large a group since the Hilbert space
would decompose into representations of $LG \times LG$
\cite{gepner-witten.wzw}; we need only one $LG$ symmetry factor. Right
now we are at the same developmental stage as equation~(4.15) of I
where we had a Lagrangian for a superparticle moving on $G$ with
symmetry group $G\times G$. Now we can exploit the full machinery
developed in that paper: in particular, we can use the horizontal
supersymmetry construction to build a supersymmetric $\sigma$-model for
a particle moving on $LG/T$.  Our construction guarantees that the
model remains supersymmetric and only admits a left action by the loop
group $LG$ as a symmetry; in projecting down from $LG$ to $LG/T$ we
lose the right action of $LG$ as a symmetry. If we write
$\L_{(\lambda,k)}$ as $\L_{(0,k)} \otimes \L_{(\lambda,0)}$ then  we
are still missing the implementation of the line bundle
$\L_{(\lambda,0)}$ over $LG/T$, a problem
we address in Section~\ref{lagrangian}.

The supersymmetric $LG/T$ model we schematically described above admits
the following maximal set of commuting operators:
\begin{itemize}
\item $\ham$: the hamiltonian which generates time translations;
\item $\mom$: the momentum which generates spatial translations;
\item $(-1)^F$: the fermion parity operator;
\item $\{H_i\}$: a basis for the Cartan subalgebra corresponding to
the left $T$ action on $LG/T$.
\end{itemize}
It is important to notice that the holomorphic sector (right moving) and
antiholomorphic sector (left moving) of the $\sigma$-model are not
identical. Our $\sigma$-model has a $(0,1)$-supersymmetry which acts only
on the left moving sector. Also, the Virasoro central extensions
$c$ and $\bar c$ of the left and right moving sector do not coincide.
It is convenient to introduce the operators
$\l$ and $\lb$ defined by:
\bea
\ham &=& (\l-{c\over 24})+(\lb-{\bar{c}\over 24})\;,\\
\mom &=& (\l-{c\over 24})-(\lb-{\bar{c}\over 24})\;.
\eea
The supersymmetry generator $\bar{G}_0$ is related to $\lb$ by
\beq
\label{G}
\bar{G}^2_0=\lb-{\bar{c}\over 24}\;.
\eeq
Of fundamental importance in our work is the quantum mechanical partition
function
\bea
Z(\theta, \tau_1, \tau_2) &=&
\Tr (-1)^F e^{i\theta} e^{2\pi i \tau_1 \mom} e^{-2\pi \tau_2
\ham} \nonumber\\
\label{partition}
&=& \Tr (-1)^F e^{i\theta} q^{\l -c/24}
	\left(\qb\right)^{\lb-\bar{c}/24}
\eea
where $q= \exp(2\pi i \tau)$, $\theta= \sum_j \theta^j H_j \in \t$, and
$\tau = \tau_1 + i \tau_2$.
Using $\{(-1)^F, \bar{G}_0\}=0$ and the usual pairing of states argument
(implied by \ref{G}) one
concludes that the full trace reduces to a trace only over the kernel of
$\bar{G}_0$. This kernel consists of precisely the supersymmetric states of
the theory, namely those states $\Psi$ of the Hilbert space which
satisfy $\bar{G}_0 \Psi = 0$. The partition function may be written as
\beq
	\label{susy-part}
Z(\theta,\tau_1,\tau_2) =\Tr_{\rm{SUSY}} (-1)^F e^{i\theta}
	q^{\l-c/24} \;.
\eeq
In the above $\displaystyle \Tr_{{\rm SUSY}}$ means the trace
only over the kernel of $\bar{G}_0$. Note that
$Z(\theta,\tau_1,\tau_2)$ is an analytic function of $\tau$ and that
it is the character index of $\bar{G}_0$. The analyticity of the
partition function in $\tau$ plays a crucial role in our path integral
computations. We will study the path integral in the $\tau_2\to 0$
limit. In this limit, the path integral is dominated by critical points
 and we show that the quadratic approximation near the critical
points leads to an analytic function of $\tau$. The corrections to the
quadratic approximation are a power series in $\sqrt\tau_2$ and thus
will not be analytic. Supersymmetry tells us that all these terms must
vanish. Thus the path integral in the $\tau_2\to 0$ limit may be used
to calculate the index.

At the risk of repeating ourselves, perhaps a more mathematical
synopsis of this paper would be useful.  We learn from examining the
elliptic genus that there are two ways of computing the $S^1$--index of
the Dirac operator $\dslash$ on $LM$ (in the weak coupling
limit~\cite{taubes.elliptic}). One can use a fixed point formula or one
can use path integrals generalizing the supersymmetric quantum
mechanics derivation of the index formula for the Dirac
operator~\cite{ag.index,friedan-windey.index,witten.asindex}.

In \cite{pressley-segal.loopgroups} one finds a heuristic sketch
deriving the Weyl-K\v{a}c character formula via the fixed point method
extending Atiyah and Bott for the Weyl character formula. As a warm
up exercise,  we derived the Weyl formula via path integrals in I. Here
we ``complete the square'' by using path integrals to obtain the
Weyl-K\v{a}c formula.

One expects that the extension from $G$ to $LG$ should be routine but
there are several obstacles. First, the standard supersymmetric
non-linear sigma model Lagrangian (\ref{NLS}) is not invariant under
left or right translation by elements of LG (because of the derivative
in the $S^1$ direction). Adding a Wess-Zumino term
restores $LG$ invariance, and has a geometric interpretation as parallel
transport for a line bundle with connection over $LG$.

Now the Lagrangian for paths on $LG/T$ is simple: the usual kinetic
term for the curve and its fermionic partner (a tangent vector field
along the loop), potential energy terms associated with the natural
vector field on $LG$,  plus a Wess-Zumino term we have just described.
Although the Lagrangian is conceptually simple, it is not amenable to
computation. We need to lift curves in $LG/T$ to curves in $LG$ which
are, of course, maps of a cylinder (or torus) into $G$. An essential
step is the lifting of supercurves on $LG/T$ to superhorizontal curves
on $LG$. For simplicity we discuss the nonsupersymmetric case (the
reader can verify by using concepts developed in
Section~\ref{lagrangian} that all the arguments we shall give go through
in the supersymmetric case). We can then express the original
lagrangian in terms of a lagrangian on the lifts. That is done locally
by a local splitting of $LG$ into $\gt(U) \times T$ using a section
$\gt: U \subset LG/T \to LG$. One is finally in a position to compute
the path integral by the steepest descent approximation at the fixed
points of the action of $T$ on $LG/T$.

We now present the geometric background in a little more detail
(see \cite[Chapter 4]{pressley-segal.loopgroups}). The
space $LG$ has a natural bi-invariant inner product which at
the identity element is the inner product on the Lie algebra of $LG$:
\beq
\IP{X}{Y} = \int_0^1 dx \; K( X(x), Y(x))\;.
\eeq
Hence $LG/T$ has an inherited inner product, and $LG$ is a principal
bundle with group $T$ and has a natural connection $\omega$ --- the
orthogonal complement of $T$-orbits.

The evaluation map $e: S^1 \times LG \to G$ gives a closed left
invariant 2--form $\Lambda$ on $LG$, given by the formula
\bea
	\Lambda &=& -2\pi i  \int_{S^1} e^* \sigma \;,\\
	\label{Lambda-def}
	&=& {1\over 2\pi i \K} \int_0^1 dx\;
	\Tr\left( g^{-1} {dg\over dx}\; g^{-1}\delta g \wedge
			g^{-1} \delta g \right)\;,
\eea
where $\sigma$ is the basic integral 3--form on $G$ generating
$H^3(G,{\Bbb Z})$. Now $i\Lambda/2\pi$ is in $H^2(LG,{\Bbb Z})$ and
so defines a line bundle $\L_\Lambda$ over $LG$ with connection whose
curvature is $\Lambda$. But $\Lambda$ is not the pull
back of a 2--form on $LG/T$. It appears as if the standard Wess-Zumino
term on $LG$ cannot be used to describe motion on $LG/T$ since it does
not descend. We will see that this is not so.

We could instead have used the 2--form $\Omega$ on $LG$ with
\beq
	\Omega(X,Y) = {i \over \pi \K}\; \IP{{dX \over dx}}{Y}\;.
\eeq
For conceptual\footnote{Stone~\cite{stone.lg} has studied the  WZW
action using the form $\Omega$ from a geometric quantization
viewpoint. Alekseev and Shatashvili~\cite{alek-shat.geom} have also
discussed loop groups and their representations from the point of view of
geometric quantization.}
use $\Omega$ is much better than $\Lambda$ because it is
left invariant under $LG$.  It is easy to see that $i\Omega/2\pi$ is
the pull back of a closed 2--form $i\Omtt/2\pi$ on $LG/T$ which is
integral so that $\cal L$ is the pullback of a line bundle $\TT{\cal
L}$ with connection $\TT{B}$.

We want to be as close to the WZW model as possible for practical
reason, \ie, we would like to use $\Lambda$. But $\Lambda = \Omega +
d\mu$ where $\mu$ is the 1--form on $LG$:
\beq
	\mu(X) = {1\over 2\pi i \K} \IP{g^{-1}\,{dg\over dx}}{X}
\eeq
at $g(x)$. Although $\mu$ does not come from $LG/T$, it is
right invariant under $T$. Split $\mu$ into $\mu_v + \mu_h$, its
vertical and horizontal pieces, so that
\beq\mu_v(W) = {1\over 2\pi i \K} \IP{g^{-1}\,{dg\over dx}}{\omega(W)}\;.
\eeq
Now $\mu_h$ is the pullback of $\TT{\mu}_h$ and we
can modify the connection $\TT{B}$ to $\TT{B} + \TT{\mu}_h$, with
curvature $\TT{\Omega} + d\TT{\mu}_h$. We use this connection in a
Wess-Zumino term; when we lift to horizontal curves, we get the same
Wess-Zumino term as using $\Lambda$ and its connection $A$. That is,
$\Lambda =dA$, $\Omega = dB$ and $\Lambda -\Omega = d(A-B) = d\mu=
d\mu_v + d\mu_h$. Hence $A-(B + \mu_h) = \mu_v + df$ for some function
$f$ since $LG$ is simply connected. The function $f$ may be absorbed
into the choice of $A$ by letting $A \to A -df$. This does not change the
curvature $\Lambda$. On horizontal lifts $\mu_v$ is zero so
\beq
	\label{A-con}
	\int_C A = \int_C (B+\mu_h) \;,
\eeq
where $C$ is the horizontal
lift of the path $\gamma$ on $LG/T$ up to the bundle.
Formula (\ref{A-con}) is very important from the practical viewpoint
because it means that we can use the Wess-Zumino
term $\int \Lambda$ on $LG$ in our calculations.

The path integral for the motion of a particle on $LG/T$
which we have to evaluate to get the Weyl-K\v{a}c
character formula is a supersymmetric variant of the following:
\beq
	\label{schematic}
	\int_{{\cal P}(\ell)} \rho(u,\ell)
	\exp\left\{	\vphantom{\biggl(}
		-I_K[\gamma(u,\ell)] -I_V[\gamma(u,\ell)]
		- k I_P[\gamma(u,\ell)]
		-I_T[\gamma(u,\ell)]\right\}\;.
\eeq
${\cal P}(\ell)$ is the set of all paths $\gamma(u,\ell)$ with initial
point $u\in LG/T$ and  endpoint $\ell\cdot u$ being the translate of
$u$ by the induced action of $\ell\in T$ on $LG/T$. The kinetic energy
contribution to the action $I_K[\gamma(u,\ell)]$ is simply the square
of the velocity integrated along the curve. The potential energy term
$I_V[\gamma(u,\ell)]$ is the  square of the natural $S^1$ vector field
on $LG/T$ (induced from the natural $S^1$ action on $LG$) integrated
along the curve. The parallel transport term, $k I_P[\gamma(u,\ell)]$,
is parallel transport on
the $k$-th power of $\L$ via the connection $k(\TT{B} + \TT{\mu}_h)$.
Finally we need to select a $T$ character and for this we use the
induced natural $T$-connection $\omega$ on an associated homogeneous
line bundle with infinitesimal $T$-character $\lambda$.
$I_T[\gamma(u,\ell)]$ is parallel transport on this line bundle. Thus
we see that we have a quantum mechanical system whose wave function is
a section of a homogeneous line bundle (with connection)
over $LG/T$ which  we shall
denote by $\L_{(\lambda,k)}$. Now $\ell\in T$ acts on this line bundle
and maps the fiber over $u$ into the fiber over $\ell\cdot u$ via a map
$\rho(u,\ell)$. Putting all this together we see that (\ref{schematic})
is gauge invariant.

It is possible to write down the full supersymmetric action on $LG/T$,
but it is cumbersome to do so. It is expressed most easily on $LG$.
%\newpage

\section{The Lagrangian}
\label{lagrangian}

Let us summarize briefly what we have done so far. At the one loop
level, the Wess-Zumino-Witten model can be seen either as a modified
$\sigma$--model on a torus with target space $G$ or the quantum
mechanics for a particle moving on $LG$, the loop group of $G$.  In the
former approach one knows that the Wess-Zumino term renders the theory
conformally invariant and that there exists an infinite number of
conservation laws corresponding to the generators of an affine Lie
algebra at level $k$ and the associated Virasoro algebra.  In the
latter approach which better corresponds to the geometrical intuition
we have tried to convey, the Wess-Zumino term corresponds exactly to a
coupling of the particle to a $U(1)$ gauge field.  This coupling,
linear in the time derivative, is of the form $A_\mu\dot{x}^\mu$ and
the gauge field comes from the $U(1)$ central extension
$\widetilde{LG}$ of the loop group.  It was explained previously why we
have to build the operator $\bar{\partial} \otimes
I_{\L_{(\lambda,k)}}$ and how it corresponds to the generator of a
chiral $(0,1)$ supersymmetry. We then built the supersymmetric
extension of this model but we still need its projection to the coset
space $\widetilde{LG}/(T\times U(1)) = LG/T$.

The trick to constructing a supersymmetric lagrangian on $LG/T$ is to
exploit the discussion of the previous section on the supersymmetric WZW
model. Let us temporarily forget about supersymmetry and review how one
would construct a bosonic lagrangian on $LG/T$ given lagrangian (\ref{NLS})
on $LG$ as a starting point. For pedagogical reasons we begin by discussing
the example of I. The lagrangian for a particle moving on $G$ is
$(g^{-1}(y) \d_y g(y))^2$, where $g(y)$ is the curve on $G$. How does one
construct the lagrangian for the motion of the particle on $G/T$? One
notices that $G$ is a principal $T$-bundle over $G/T$ with a bi-invariant
metric and a natural $T$ connection $(g^{-1} dg)_{\t}$. Thus $G/T$ has a
natural metric $\langle\cdot,\cdot\rangle$ induced by the horizontal spaces
of the $T$-connection. A curve $u(y)$ on $G/T$ has a unique horizontal lift
to $G$ (after specifying the starting point) which we will call $\ghor(y)$.
{}From the geometry it is clear that the natural lagrangian on $G/T$:
$\langle \d_y u(y),
\d_y u(y) \rangle$ is the same as
\beq
\label{G-lag}
\Tr\left(\ghor^{-1}(y) \d_y\ghor(y)\right)^2\;.
\eeq
We remark that the right invariance of (\ref{G-lag}) under the action of
$T$ shows that (\ref{G-lag}) is independent of the starting point for the
lift. This invariant description suffers at the practical level. Namely,
$\ghor(y)$ is a complicated solution to a differential equation and thus
$\ghor(y)$ is not very useful in a path integral computation. The solution
to our dilemma is to give a local reformulation of the invariant
description by exploiting the principal $T$-bundle structure in such a way
that everything will patch smoothly. Let $\gt:G/T \to G$ be a local
section. We can lift the curve $u(y)$ on $G/T$ to $G$ as $\gt(u(y))$. We
know that $\ghor(y)$ and $\gt(u(y))$ are related by an element of $T$:
$\ghor(y) = \gt(u(y)) t^{-1}(y)$. By using the $T$-connection we see that
locally
\beq
\Tr\left(\ghor^{-1}(y) \d_y\ghor(y)\right)^2 =
\Tr\left(\gt^{-1}(u(y)) \d_y \gt(u(y))\right)^2_{\m}\;.
\eeq
We leave it as an exercise to the reader to verify that the local
description patches together in a natural way.
Thus a section can be used to locally describe the Lagrangian in a way
which as we shall see
is amenable for efficient path integral use. For example, a
convenient section near the identity of $G$ is to write $\gt = \exp
\phit_{\m}$ where $\phit_{\m}$ has values in $\m$, and a convenient
section near any other point is the left translate of $\exp \phit_{\m}$.

Let us introduce some notation for discussing the $LG/T$ case.   An
element in $LG$ will be written $g(x)$, $x\in [0,1]$, and an element in
$T$ will simply be written $t$.  Curves on these spaces
will also depend on the time variable $y\in [0,\tau_2]$.  From a
two dimensional viewpoint we will have fields $g(x,y)$ and $t(y)$
together with their respective supersymmetric partners\footnote{Please
note that the modular parameter of the torus is denoted by $\tau$ while
the supersymmetric partner of $t$ is denoted by $\th$.} $\gamma(x,y)$
and $\th(y)$.  The variables $(x,y)$ parametrize the two dimensional
torus. We also define for later use the complex variables $z=x+iy$ and
$\bar{z}=x-iy$.  The generator of supersymmetry is given by ${\bf
Q}=\partial_\theta-\theta\partial_{\bar{z}}$, where $\theta$ is a
grassmann variable of weight $(0,-\frac{1}{2})$. We use $\delta$ to
denote the differential on the infinite dimensional space of fields.

To commence our discussion of the $LG/T$ case we forget about
supersymmetry. The nonlinear sigma model (\ref{NLS}) describes the
evolution of a curve $g(x,y)$ in $LG$. This lagrangian has both a kinetic
energy term
$$\int dx \Tr( g^{-1}(x,y) \d_y g(x,y))^2$$
and a potential energy term
$$\int dx \Tr( g^{-1}(x,y) \d_x g(x,y))^2 \; .$$
To construct a natural lagrangian on $LG/T$ induced from (\ref{NLS}) we
exploit that $LG$ is a principal $T$-bundle over $LG/T$ with a bi-invariant
metric and a natural $T$-connection. The $T$--connection on the bundle is
defined as follows. The connection $1$-form $\omega$ maps a tangent vector
to $LG$ at $g(x)$ into an element of $\t$. The tangent vector translated to
the identity in $LG$ is an element of the Lie algebra of $LG$, namely
$L\g$, and denoted by $X(x)$. Project for each $x$ onto $\t$ and integrate
over $S^1$:
\beq
\omega(X) = \int_{S^1}dx\;X(x)_{\t}\;.
\eeq
In terms of the
left invariant differential forms on $LG$ this may be written as
\beq
	\label{T-con}
	\omega = \int_0^1 dx\; (g^{-1}(x)\delta g(x))_{\t}\;.
\eeq
More geometrically, $\omega$ is orthogonal projection of the tangent
space to $LG$ onto the tangent space to the orbit of $T$ relative to the
bi-invariant metric on $LG$ which at the identity is
$\int_{S^1} K(\cdot\, ,\cdot)$. It follows that $LG/T$ has a natural
metric $\langle \cdot, \cdot \rangle$ induced by the horizontal
spaces of the connection. A curve $u(y)$ on $LG/T$ has a unique
horizontal lift (after specifying the initial point) to $LG$ which we
will call $\ghor(x,y)$. From the geometry it is clear that the natural
kinetic energy term on $LG/T$: $\langle \d_y u(y), \d_y u(y)\rangle$
may be written as
\beq
	\int_0^1 dx\; \Tr\left( \ghor^{-1}(x,y) \d_y \ghor(x,y) \right)^2\;.
\eeq
Note that the potential energy term on $LG$ descends to a function
$V[u]$ on $LG/T$ which is defined by
\beq
	V[u] = \int_0^1 dx\; \Tr \left(\ghor^{-1}(x,y) \d_x \ghor(x,y)
	\right)^2\;.
\eeq
We find ourselves in much the same situations as discussed in the $G/T$
case. Although we have an invariant formulation it turns out that working
with $\ghor$ is impractical. We give a local description which patches
together nicely. Let $\gt:LG/T \to LG$ be a local section. We can lift the
curve $u(y)$ on $LG/T$ to $LG$ as $\gt(u(y))$. We know that $\ghor(x,y)$
and $\gt(u(y))$ are related by an element of $T$: $\ghor(x,y) = \gt(u(y))
t^{-1}(y)$. By using the connection $\omega$ we see that the horizontal
condition on $\ghor(x,y)$ requires $t(y)$ to satisfy the differential
equation
\beq
	\label{hor0}
	\int_0^1 dx\; \left( \gt^{-1}(u(y)) \d_y \gt(u(y))\right)_{\t}
		- \d_y t(y)\; t^{-1}(y) = 0 \;.
\eeq
If we define Fourier modes
\beq
	\left( \gt^{-1}(u(y)) \d_y \gt(u(y))\right)_{\t} =
\sum_{n=-\infty}^\infty \H_{y,n}(y) e^{2\pi i nx}
\eeq
then one can see that the kinetic energy term may be written as
\beq
	\int_0^1 dx\; \Tr \left( \gt^{-1}(x,y) \d_y \gt(x,y) \right)^2_{\m}
		+ \sum_{n \neq 0} \Tr \H_{y,n}(y) \H_{y,-n}(y)\;.
\eeq
One can verify that the kinetic energy term above patches nicely.
We leave the potential energy term as an exercise to the reader.

We now return to the supersymmetric discussion associated with the
$LG/T$ case.  In what follows we will often suppress the coordinate
dependence of the fields but it is important to remember that since $t$
and $\th$ belong to $T$ and its tangent space and not to $LT$, they do
not depend on the spatial coordinate $x$.

We will now use the natural $T$-connection (\ref{T-con}) on
$LG$ to induce supersymmetry on $LG/T$ from a naturally formulated
supersymmetry on $LG$. This is precisely analogous to using a
connection to define the horizontal tangent spaces on the bundle and
relating these to tangent spaces on the base. The importance of our
construction is that it allows us to express the supersymmetric model
on $LG/T$ in terms of quantities defined on $LG$ suitable for path
integral use. Firstly we must define supersymmetry on $LG$. Consider a
supercurve which may be expressed in superfield notation as ${\bf
G}(x,y) = g(x,y) e^{\theta\gamma(x,y)}$ (see~I). The supersymmetric
variation of  ${\bf G}$ is given by
\beq
\delta_s{\bf G}=\varepsilon{\bf Q}{\bf G}\;,
\eeq
where $\varepsilon$ is the anticommuting parameter of the transformation.
In terms of components the supersymmetry transformations are given by
\bea
\label{susy1}
g^{-1}\delta_sg &=& \varepsilon\gamma\;,\\
\label{susy2}
\delta_s\gamma &=& \varepsilon\left(g^{-1}\partial_{\bar{z}} g-\gamma
    \gamma\right)\;.
\eea
Note that the supersymmetry transformations are equivariant under the
right $T$-action on $LG$.

How do we lift a supercurve on $LG/T$ to a {\it
superhorizontal\/} curve on $LG$?
The condition that a curve in $LG$ is the
horizontal lift of a curve in $LG/T$ is that the global 1--form
$\omega=\int^1_0 dx\left(g^{-1}\delta g\right)_{\t}$ vanish when evaluated
along the curve.  We generalize this to the supersymmetric case by
noticing that one can interpret (\ref{susy1}) as a tangent vector; thus
it is natural to impose  the vanishing of
$\omega_s=\int^1_0 dx\left(g^{-1}\delta_s g\right)_{\t}$
as the first superhorizontal condition.
Using (\ref{susy1}) we see that this condition is simply
\beq
	\label{ghoriz1}
		\int_0^1 dx\; \gamma_{\t}(x,y) = 0\;.
\eeq
For consistency we must also impose that the
supersymmetric transform of (\ref{ghoriz1}) also vanish:
\beq
	\label{ghoriz2}
	\int_0^1 dx\; \left(g^{-1}(x,y)\partial_{\bar{z}} g(x,y)
	-\gamma(x,y)\gamma(x,y)\right)_{\t} = 0\;.
\eeq
If one forgets about the fermions then the above is almost the condition
that the lift be horizontal in the ordinary sense\footnote{It would be the
standard condition if it was a derivative with respect to $y$, see
(\ref{hor0}).}. The additional term is a Pauli type coupling (see~I). Note
that the formulation of superhorizontal has been done in a global way.

The equivariance of the superhorizontality conditions tells us that
arguments concerning the lagrangian we gave in the bosonic case will go
through in the supersymmetric case. For example, the supersymmetric
kinetic energy term on $LG/T$ may be formulated on $LG$ by the use of
superhorizontal lifts.

We now turn to the local parametrization of supersymmetry and
superhorizontal lifts.  We parametrize a  loop $g(x)$ in $LG$ by a
local section $\widetilde{g}$ and an element of $T$ as
$g(x)=\widetilde{g} t^{-1}$.  Using the decomposition of the Lie
algebra of $G$ into  ${\euf g}={\euf t}\oplus{\euf m}$ leads to the
equations
\bea \left(g^{-1}\delta
g\right)_{\t} &=& \left(\widetilde{g}^{-1}
     \delta\widetilde{g}\right)_{\t} -dtt^{-1}\;,\\
\left(g^{-1}\delta g\right)_{\m} &=& t\left(\widetilde{g}^{-1}
      \delta\widetilde{g}\right)_{\m} t^{-1} \;.
\eea
To  find the equivalent relations for $\gamma$ it is
best to reintroduce a superfield notation ${\bf
G}=ge^{\theta\gamma}$.
We have the local parametrization
${\bf G}={\bf\widetilde{G}}{\bf T}^{-1}$ given by the local
supersections ${\bf \widetilde{G}} = \gt e^{\theta \widetilde{\gamma}}$
and superfiber variables ${\bf T}=te^{\theta\th}$. This  gives
\bea
\label{paran1}
\gamma_{\t} &=& (\widetilde{\gamma})_{\t}-\th \;, \\
\label{paran2}
\gamma_{\m} &=& (t\widetilde{\gamma}t^{-1})_{\m}\;.
\eea

In terms of the section, the $T$-connection in local coordinates
 may be written as $\omega=A-dtt^{-1}$ where
\beq
\label{T-connection}
	A=\int^1_0 dx\;\left(\widetilde{g}^{-1}(x)\delta
		\widetilde{g}(x)\right)_{\t}\;.
\eeq
This is the connection we will use to get local formulas.

The supersymmetry transformations of the fiber $T$ are given by
(defining $t = \exp f$)
\bea
\label{susy3}
\delta_stt^{-1} &=& \delta_s f =\varepsilon\th\;,\\
\label{susy4}
\delta_s\th &=& \varepsilon\partial_{\bar{z}} f \;.
\eea
We now have enough information to formulate the supersymmetry
transformations of the local sections:
\bea
\left(\widetilde{g}^{-1}\delta_s\widetilde{g}\right)_{\m}(x,y)&=&
   \varepsilon\widetilde{\gamma}_{\m}(x,y)\;,\\
\left(\widetilde{g}^{-1}\delta_s\widetilde{g}\right)_{\t}(x,y)&=&
    \delta_s f  (y)+\varepsilon\left(\widetilde{\gamma}_{\t}(x,y)
    -\th(y)\right)\nonumber\\
&=& \varepsilon\widetilde{\gamma}_{\t}(x,y)\;.
\eea
In the last equation we have used (\ref{susy3}).

Using similar algebraic manipulations will give the supersymmetric
transformation of the fermionic partner of  $\widetilde{g}$.
Expressing $\gamma$ in terms of the section we have
\bea
\delta_s\gamma &=& \delta_s\left(t\widetilde{\gamma}t^{-1}-\th
	\right)\nonumber\\
 &=& \delta_s t\widetilde{\gamma}t^{-1}+t\delta_s\widetilde{\gamma}t^{-1}
      -t\widetilde{\gamma}t^{-1}\delta_stt^{-1}-\delta_s\th\nonumber\\
 &=&\varepsilon\th t\widetilde{\gamma}t^{-1}
	+\varepsilon t\delta_s\widetilde{\gamma}
       t^{-1}+\varepsilon t\widetilde{\gamma}t^{-1}\th-\varepsilon
       \partial_{\bar{z}}tt^{-1}\;.
\eea
The same variation can be written by the use of  (\ref{susy2})
which in terms of the section reads
\beq
\delta_s\gamma=\varepsilon t\left[\widetilde{g}^{-1}\partial_{\bar{z}}
   \widetilde{g}-t^{-1}\partial_{\bar{z}}t-\widetilde{\gamma}\widetilde
   {\gamma}+\widetilde{\gamma}\th+\th\widetilde{\gamma}\right]t^{-1}
	\;.
\eeq
Comparing the two expressions we find
\beq
\delta_s\widetilde{\gamma}=\varepsilon\left(\widetilde{g}^{-1}\partial
    _{\bar{z}}\widetilde{g}-\widetilde{\gamma}\widetilde{\gamma}\right)
\eeq
which can be decomposed as
\bea
\delta_s\widetilde{\gamma}_{\m} &=& \varepsilon\left((\widetilde{g}^{-1}
     \partial_{\bar{z}}\widetilde{g})_{\m}-(\widetilde{\gamma}\widetilde
     {\gamma})_{\m}\right)\;,\nonumber\\
\delta_s\widetilde{\gamma}_{\t} &=& \varepsilon\left((\widetilde{g}^{-1}
     \partial_{\bar{z}}\widetilde{g})_{\t}-(\widetilde{\gamma}_{\m}\widetilde
     {\gamma}_{\m})_{\t}\right)\;.
\eea

We are now ready to express the superhorizontality conditions in terms
of the local section. We denote the superhorizontal lifts of the
supercurve on $LG/T$ by $\ghor$ and $\gamhor$.
{}From (\ref{paran1}) we find the first condition
\beq
\label{horiz1}
0=\int^1_0 dx\left(\widetilde{\gamma}_{\t}(x,y)-\th(y)\right)\;.
\eeq
Applying a
supersymmetry transformation to this equation we find the second
horizontality condition
\bea
0 &=& \int^1_0 dx\left(\delta_s\widetilde{\gamma}_{\t}(x,y)-\delta_s\th
    (y)\right) \nonumber \\
\label{horiz2}
  &=&  \int^1_0 dx\left[(\widetilde{g}^{-1}\partial_{\bar{z}}\widetilde
  {g})_{\t}-(\widetilde{\gamma}_{\m}\widetilde{\gamma}_{\m})_{\t}\right]
  -\partial_{\bar{z}}tt^{-1}\;.
\eea
Using the mode expansion
\beq
\label{modes}
\widetilde{\gamma}_{\t}(x,y)
=\sum^\infty_{n=-\infty}\widetilde{\gamma}_{\t,n}(y)   e^{2\pi inx}\;,
\eeq
the first condition (\ref{horiz1}) gives
\beq
\label{horiz1.mode}
\widetilde{\gamma}_{\t,0}-\th=0 \;.
\eeq
{}From the expressions (\ref{paran1}) and (\ref{paran2}) we find
$\gamhor=t\widetilde{\gamma}_{\m} t^{-1}+\widetilde{\gamma}_{\t}-\th$.
Equivalently, using (\ref{modes}) and (\ref{horiz1.mode}), the
final  form of the  superhorizontal lift is
\beq
\gamhor(x,y)=t(y)\widetilde{\gamma}_{\m}(x,y)t^{-1}(y)+\sum_{n\neq 0}
     \widetilde{\gamma}_{\t,n}(y)e^{2\pi inx}\; .
\eeq
 Note the important fact that the absence of zero modes implies that all
dependence on $\th$ has disappeared.  Similar algebraic manipulations
and the mode expansion
\beq
\label{g-modes}
\left(\widetilde{g}^{-1}\partial_a\widetilde{g}\right)_{\t}=
    \sum^\infty_{n=-\infty}\H_{a,n}(y)e^{2\pi inx}
	\qquad\hbox{for $(a=z,\bar{z})$}.
\eeq
give
\bea
n\neq 0:\qquad \left(\ghor^{-1}\partial_a \ghor\right)_{\t,n}&=&\H_{a,n}\;,\\
n=0:\qquad\left(\ghor^{-1}\partial_a \ghor\right)_{\t,0}
	&=&\H_{a,0}-\partial_a f \;.
\eea
{}From (\ref{T-connection}) we see that the $T$-connection in local
coordinates is given by
\beq
	A(y) = \int_0^1 dx\; \H_y(x,y) = \H_{y,0}(y)
\eeq
Note that $\H_{x,0}$ is gauge invariant with respect to $T$~gauge
transformations.

It is now a matter of algebra to project the kinetic part of the
SUSY-WZW lagrangian on $LG$ to $LG/T$:
\bea
\int d^2z \Tr\left(\ghor^{-1}\partial_z\ghor
	\ghor^{-1}\partial_{\bar{z}}\ghor\right)
    &+&\int d^2z\Tr\gamhor\partial_z\gamhor\nonumber\\
&=&\int d^2z\Tr\left(\widetilde{g}^{-1}\partial_z\widetilde{g}\right)_{\m}
    \left(\widetilde{g}^{-1}\partial_{\bar{z}}\widetilde{g}\right)_{\m}
    \nonumber\\
&+&\sum_{n\neq 0}\int dy\Tr\H_{z,n}\H_{\bar{z},-n}\nonumber\\
&+&\int dy\Tr(\H_{z,0}-\partial_z f )(\H_{\bar{z},0}-\partial_
    {\bar{z}} f )\nonumber\\
&+&\int d^2z\;\Tr\widetilde{\gamma}_{\m} \left(
	\partial_z\widetilde{\gamma}_{\m}
     + [\partial_z f ,\widetilde{\gamma}_{\m}]\right) \nonumber\\
&+&\int d^2z\;\xi_{\t}\partial_z\xi_{\t} \;.
\eea
In the above, $\xi_{\t}$ is defined by
\beq
\xi_{\t}=\sum_{n\neq 0}\widetilde{\gamma}_{\t,n}(y)e^{2\pi inx}\;.
\eeq
The Wess-Zumino term on $LG$ restricted to a superhorizontal
lift becomes
\beq
  \int \Tr\left(\ghor^{-1}d\ghor\right)^3 =
	\int \Tr \left(\gt^{-1} d\gt\right)^3
	-6i\int dy\Tr
\left(\H_{z,0}\partial_{\bar{z}} f -\H_{\bar{z},0}\partial_z f \right)\;.
\eeq

For completeness we list below the expression of the superhorizontal
lift in terms of the local section for each term in the WZW action.
The bosonic kinetic energy term is given by
\bea
	\int d^2z\; \Tr \left(\ghor^{-1}\partial_z\ghor
	\ghor^{-1}\partial_{\bar{z}}\ghor\right)
&=&\int d^2z\Tr\left(\widetilde{g}^{-1}\partial_z\widetilde{g}\right)_{\m}
    \left(\widetilde{g}^{-1}\partial_{\bar{z}}\widetilde{g}\right)_{\m}
    \nonumber\\
&+& \sum_{n\neq 0} \int dy\; \Tr \H_{n,z} \H_{-n,\zb}
	\nonumber\\
&+& \int dy\; \Tr \H_{x,0} \left(\gamt_{\m}\gamt_{\m}\right)_{\t,0}
	\nonumber\\
&-& \int dy\; \Tr	\left(\gamt_{\m}\gamt_{\m}\right)_{\t,0}^2 \;.
\eea
The fermionic kinetic energy term is
\bea
	\int d^2 z\; \Tr \gamhor \d_z \gamhor &=&
	\int d^2 z\; \Tr \xi_{\t} \d_z \xi_{\t}
	+ \half \int d^2 z\; \Tr \gamt_{\m} \d_x \gamt_{\m}
		\nonumber\\
	&-& {i\over 2} \int d^2 z\; \Tr \gamt_{\m}
	 \left( \d_y \gamt_{\m} + \left[ A, \gamt_{\m}\right] \right)
		\nonumber\\
	&+& \int dy\; \Tr \H_{x,0} \left(\gamt_{\m}\gamt_{\m}\right)_{\t,0}
		-2 \int dy\;\Tr \left(\gamt_{\m}\gamt_{\m}\right)_{\t,0}^2
\eea
The WZ term is given by
\bea
	\int \Tr(\ghor^{-1}d\ghor)^3 &=& \left\{ \int \Tr(\gt^{-1} d\gt)^3
		+ 3 \int dy \Tr A \H_{x,0} \right\}
	\nonumber\\
	&-& 3i \int dy\; \Tr \H_{x,0}^2
	+ 6i \int dy\; \Tr \H_{x,0}\left(\gamt_{\m}\gamt_{\m}\right)_{\t,0}
\eea
One can verify that the quantity in curly braces is $T$-gauge invariant.

Collating all the terms together we find the following action for the
supersymmetric model on $LG/T$:
\bea
{i \pi \K \over 2k}\; I_{LG/T}
	&=&- \frac{i}{2}\int d^2z\Tr(\widetilde{g}^{-1}\partial_z
   \widetilde{g})_{\m}(\widetilde{g}^{-1}\partial_{\bar{z}}\widetilde{g})_{\m}
		\nonumber \\
&+&\frac{i}{2}\int d^2z \Tr\widetilde{\gamma}_{\m}\left(
	\partial_z\widetilde{\gamma}_{\m}
	+ \left[ \H_{z,0}, \gamt_{\m} \right] \right)
		\nonumber\\
   &+& \frac{1}{12}\left\{ \int\Tr(\widetilde{g}^{-1}d\gt)^3
	+ 3 \int dy\; \Tr A \H_{x,0} \right\}
		\nonumber\\
&-& \frac{i}{2}\sum_{n\neq 0}\int dy\Tr\H_{z,n}\H_{\bar{z},-n}
	+\frac{i}{2}\int d^2 z\;\Tr \xi_{\t}\partial_z\xi_{\t}
		\nonumber\\
&-&  \frac{i}{4}\int dy\Tr \H_{x,0}\H_{x,0}
		\nonumber\\
&+& i \int dy\Tr \H_{x,0}
	\left(\widetilde{\gamma}_{\m}\widetilde{\gamma}_{\m}\right)_{\t,0}
		\nonumber\\
\label{lgt-action}
&-&\frac{i}{2}\int dy\Tr\left(\widetilde{\gamma}_{\m}
	\widetilde{\gamma}_{\m}\right)_{\t,0}
	\left(\widetilde{\gamma}_{\m}\widetilde{\gamma}_{\m}\right)_{\t,0}
		\;.
\eea
It is not clear to us which is the best way of writing the above. The
reason is that the $LG/T$ lagrangian is not Lorentz invariant and therefore
there is no obvious way to group the terms. We decided on the above
grouping because it makes certain features clear. The first two lines are
the kinetic energy terms for bosons and fermions on $L(G/T)$. The third
line is the Wess-Zumino term on $LG/T$. The fourth line is the kinetic
energy terms for bosons and fermions on $LT/T$. The last three lines are
respectively $T$~gauge invariant potential energy, Yukawa and curvature
terms which are required by supersymmetry.

It is important to observe that the lagrangian we just derived solves one
of our main problems which was to find a good way of handling curves
on $LG/T$. The main tool used in this respect was the implementation
of horizontal supersymmetry which allows us to lift supercurves on
$LG/T$ to superhorizontal curves on $LG$, a space which is more
amenable to field theoretic methods.

Our next task is to describe the appropriate modifications of this basic
lagrangian which will give the coupling to the different irreducible
representations of $LG$. In what follows we will often refer to it as the
matter coupling, adopting the traditional field theoretic language. Firstly
we construct the line bundles over $LG/T$ and study the $U(1)$ and $T$
action on them. We have explained above why these bundles play such a
crucial role in the construction of the irreducible representations of
$LG$. A line bundle $\L_{(\nu,0)}$ over $LG/T$ is determined by an
appropriate one dimensional representation of the group $T$ with
infinitesimal character $\nu$. A section of $\L_{(\nu,0)}$ is the same as a
function on the entire principal bundle $\pi:LG\rightarrow LG/T$
satisfying
\beq
F(gt^{-1})=\rho(t)F(g)\ ,
\eeq
where $\rho(t)$ is the irreducible representation of $T$ with infinitesimal
character $\nu$. We will often indicate explicitly the dependence on the
point $u$ in the base $LG/T$; for example $\widetilde{g}(u)$ will stand for
a given local section of the bundle $LG$.

It is important to keep in mind the difference between the right and the
left action of $T$. The right action lets us move up and down the fiber and
tells us that the function $F$ transforms under the representation given by
$\rho$. A local section $\widetilde{F}$ of $\L_{(\nu,0)}$ must then be
parametrized by the coordinates $u$ of $LG/T$ and is defined by
\beq
\wt{F}\u=F(\gt\xu)\,.
\eeq
This determines
the left action of $T$ on $\widetilde{F}$:
\beq
L_\ell\wt{F}\u=F(\ell\tgu)\, ,
\eeq
where $\ell\in T$.
Note that $\ell\tgu$ is a new element of $LG$ in the fiber above the
point $\ell\cdot u=\pi(\ell\tgu)$ in the base.  We can use right
multiplication to relate $\ell\tgu$ to the section by introducing
$t(u,\ell)\in T$ as follows $\ell\widetilde{g}\xu =
\widetilde{g}(\ell\cdot u)\;t^{-1}(u,\ell)$ We now can rewrite the left
$T$-action on a section $\widetilde{F}$ as follows:
\bea
L_\ell\wt{F}\u &=&
	F\left(\wt{g}(\ell\cdot u)t^{-1}(u,\ell)\right)\nonumber\\
 &=& \rho(t(u,\ell))F\left(\gt(\ell\cdot u)\right)\nonumber\\
\label{left2}
&=& \rho(t(u,\ell))\wt{F}(\ell\cdot u)\;.
\eea
We see that the transformation law (\ref{left2}) for the sections has both
an orbital part and a ``spin'' part. Because of the presence of the spin
part there are some phases which will have to be accounted for in the path
integral computation.

We now have all the necessary ingredients to construct the matter
coupling at the lagrangian level.  This will be done by using a
minimal coupling scheme, \ie, by writing a term of the form $q\int dt
A_\mu\dot{x}^\mu$. Notice that such an abelian coupling term in the
lagrangian keeps track of the change in angle along the path between
the initial and final point multiplied by the appropriate ``charge''.
This fixes its normalization uniquely. In the case at hand we just
have to compute the $T$-connection which is determined by
$(\partial_{\bar{z}}t)t^{-1}$. Notice that $t=t(y)$ depends only on
the ``time'' variable $y$. We write $t = \exp f$, and with the help of
(\ref{g-modes}), rewrite the usual $T$-connection in Eq.
(\ref{T-connection}) as $A(y)=\int_0^1 dx {\cal H}_y(x,y)$. Notice
that $A$ does not depend on $x$ since this is a $T$ connection and not
an $LT$ connection.  Using this and the mode expansion introduced in
(\ref{modes}), we rewrite the supersymmetric horizontality condition
(\ref{horiz2}):
\beq
 \partial_{y}f=A(y)-i\int_0^1 dx \left[{\cal H}_{x,0} +2i \int_0^1 dx
(\widetilde{\gamma}_{\m}\widetilde{\gamma}_{\m})_{\t}\right] \, .
\eeq
Notice that the last two terms are specifically supersymmetric contributions.
This equation gives us the required
matter action:
\beq
I_T=2i\int d^2z \;
            \nu\left({\cal H}_{\bar{z}}-
(\widetilde{\gamma}_{\m}\widetilde{\gamma}_{\m})_{\t}\right)
	\;.
\eeq
This last equation exhibits the $(0,1)$ nature of the coupling as required
by our chiral supersymmetry. This action gives the desired modification of
the generator of sypersymmetry. In other words it produces the appropriate
twisting of the Dirac operator by the holomorphic $T$-bundle associated
with the infinitesimal $T$-character $\nu$ as required by the Borel-Weil
construction.

\section{The Weyl-K\v{a}c Character Formula}
\label{computation}
\subsection{The Quadratic Approximation Around the Fixed Points}

The partition function (\ref{partition}) may be computed in the
$\tau_2 \to 0$ limit as was mentioned at the end of
Section~\ref{review}. We remind the reader that the partition function
is actually an analytic function of $q = \exp(2\pi i \tau)$ and we can
exploit this fact to compute it exactly. In this limit, any time
dependent\footnote{The reader is reminded that $y$ plays the role of
time.} field configuration will lead to an action that behaves as
$1/\tau_2$ and thus such configurations will be suppressed. The
dominant contributions to the path integral will arise from static
field configurations which satisfy the appropriate boundary
conditions. Supersymmetry requires the fields to be periodic under
$z\to z+1$. However, it is clear from (\ref{partition}) and the
discussion in I that we must use twisted boundary conditions on the
fields in the time direction.  On the bosonic fields $\gt(x,y)$, the
appropriate twisted boundary conditions are given by
\beq
\label{BC}
	\gt(x+\tau_1,\tau_2)T = \ell \gt(x,0)T
\eeq
where $\ell = \exp(i\theta)$
is the element of $T$ in expression (\ref{partition}) and $\tau_1$
enters because of the rotation induced by the momentum operator $P_1$.
We now use the fact that the dominant configurations must be static and
conclude that the saddle points are described by
\beq
	\gt(x+\tau_1,0)T = \ell \gt(x,0)T \;.
\eeq
We remark that in the above, two elements of $LG$ are identified if
they differ by an element of $T$ and a translation in $x$.  This
equation has to be true for all $\ell\in T$ and thus the above may be
rewritten in the equivalent form
\beq
	\gt(x+\tau_1,0)^{-1} \ell \gt(x,0) \in T \;.
\eeq
The solution to the above is the {\it group\/}
\beq
	{\cal N} = \left\{ n \exp\left(2\pi i \cmu x\right) \mid
	 n\in N(T:G),\; \cmu \in \check{T}\right\}\;,
\eeq
where $N(T:G)$ is the normalizer of $T$ in $G$, and $\check{T}$ is the
coroot lattice (see the Appendix for a definition). If we observe that the
momentum operator $P_1$ generates a circle group $S^1$ of symmetries
of the lagrangian by translating the loop parameter then we can recast
$\cal N$ in a more group theoretical setting.  Consider the group $S^1
\ltimes LG $ and note that the maximally commuting subgroup is
$S^1 \times T$ where $S^1$ is the circle group associated with
translating the loop parameter. Our collection ${\cal N}$ is actually
the {\it group\/} $N(S^1 \times T:S^1 \ltimes LG)$ and the
quotient $\waff = {\cal N}/(S^1 \times T)$
is a group called the {\it affine Weyl group\/}
of $LG$.  From the definition we see that the affine Weyl
group $\waff$ is the semidirect product of $W$,
the ordinary Weyl group of $G$, and
$\check{T}$, the coroot lattice. Its elements are
\beq
\ewaff=(w, e^{2\pi i\cmu x})\; \mbox{with} \; w\in W\; .
\eeq
Each element of $\waff$ is associated with a fixed point of the $S^1 \times
T$ action in $LG/T$. The evaluation of the path integral by steepest
descent will require a sum over the infinite set of Weyl points described
above.

Since we only have to study the path integral in the $\tau_2 \to 0$ limit,
it is clear from the above discussion  that it will suffice to consider
the fluctuations around the Weyl points. A supersection of $LG$ near the
 Weyl point represented
by $ n \exp(2\pi i\cmu x)$ may be written as
\beq
	\wt{\bf G}(x,y) = n e^{2\pi i\cmu x} e^{\phit(x,y)}
		e^{\gamt(x,y)\theta}\;,
\eeq
where $\phit$ and $\gamt$ parametrize the fluctuations.  The superfield
is periodic under $z\to z+1$ and under $z\to z+\tau$  it
satisfies\footnote{A more detailed explanation can be found in
Section~\ref{group-action}.}
\beq
\label{bounda1}
	\wt{\bf G}(x+\tau_1,\tau_2)T = \ell \wt{\bf G}(x,0)T\;.
\eeq
If we define
\beq
	\kappa = e^{-2\pi i\cmu \tau_1} (n^{-1}\ell n) \in T
\eeq
then the boundary conditions (\ref{bounda1}) may be formulated as
\bea
	\phit(x+\tau_1,\tau_2) &=& \kappa \phit(x,0) \kappa^{-1}\;,\\
	\gamt(x+\tau_1,\tau_2) &=& \kappa \gamt(x,0) \kappa^{-1}\;.
\eea
It is important to remember that, since we are working on the coset
space $LG/T$, there is no ``translationally invariant''
mode\footnote{Note that $\int_0^1 dx\; \phit_{\t}(x,y) =0$ and
similarly for $\gamt_{\t}$.} in $\phit_{\t}$ and $\gamt_{\t}$.
We also note that $\ell_w \equiv n^{-1}\ell n$ only depends
on the choice of coset $w=nT$ (see I). We shall often, by abuse of
notation, write $w^{-1} \ell w$ to remind the reader that it only
depends on the choice of an element of the Weyl group of $G$. The
above boundary conditions may be equivalently written as
\bea
\phit_{\m}(x+\tau_1,\tau_2) &=& \kappa \phit_{\m}(x,0) \kappa^{-1}\;,\\
\gamt_{\m}(x+\tau_1,\tau_2) &=& \kappa \gamt_{\m}(x,0) \kappa^{-1}\;,\\
\phit_{\t}(x+\tau_1,\tau_2) &=&  \phit_{\t}(x,0) \;,\\
\gamt_{\t}(x+\tau_1,\tau_2) &=&  \gamt_{\t}(x,0)  \;.
\eea

It is easy to verify that to quadratic order near a Weyl point we have:
\bea
	A(y) &=& -\;{1\over 2} \int_0^1 dx\;
		\left[\phit_{\m},\d_y\phit_{\m}\right]_{\t} + O(\phit^3)
		\;, \\
	{\cal H}_{x,0} &=& 2\pi i \cmu + {1\over 2}\;
	\int_0^1 dx\;  \left[ \d_x \phit_{\m} +
	\left[ 2\pi i \cmu, \phit_{\m}\right], \phit_{\m}\right]_{\t}
		+ O(\phit^3)\; .
\eea
These expressions enter in the perturbative expansion of the full
lagrangian near the fixed points. After a considerable amount of algebra
one finds the following relatively simple form for the lagrangian to
quadratic order near a fixed point:
\bea
	I_{\rm total}^{(2)} &=& {k\over  4\pi\cox}\; \left\{ \;
		2 \pi^2\tau_2 \Tr(\cmu\cmu)
		-{8\pi^2\cox\over k}\;\tau_2 \Tr(\nu\cmu)
		\right. \nonumber\\
&-& \int d^2 z\;\Tr\left(\d_z \phit_{\m}- {1\over2}\;
	\left[ 2\pi i\cmu - {8\pi i\cox\over k}\;\nu, \phit_{\m}
		\right] \right)
	\left(\d_{\zb}\phit_{\m} + {1\over 2}\;
			\vphantom{8\pi i\cox\over k}
		\left[  2\pi  i \cmu, \phit_{\m}\right] \right)\nonumber\\
&+& \int d^2 z\; \Tr \gamt_{\m} \left(\d_z \gamt_{\m}- {1\over2}\;
	\left[ 2\pi i\cmu - {8\pi i\cox\over k}\;\nu, \gamt_{\m}
		\right] \right) \nonumber\\
&-& \int d^2 z\; \Tr \d_z \phit_{\t} \d_{\zb}  \phit_{\t}\nonumber\\
\label{quad-action}
&+& \left. \int d^2 z\; \Tr \gamt_{\t} \d_z  \gamt_{\t} \;
	\vphantom{8\pi^2\cox\over k} \right\} \; .
\eea
Formula~(\ref{quad-action}) is the final form of the quadratic
part of the lagrangian we will use; its
derivation is nontrivial and involves subtle and delicate
cancellations which reflect the underlying geometry. The
bosonic degrees of freedom which appear in the above are a
representation of the following geometrical fact.  There is a local
equivalence between $LG/T$ and the space
\beq
	{LG \over LT} \times {LT \over T}\;.
\eeq
More precisely, $LG/T$ is a principal $LT/T$ bundle over $LG/LT$.
Notice that $LG/LT$ is the configuration space for an ``ordinary''
sigma model since one can also show that $LG/LT = L(G/T)$. Locally, the
fields in our model may be thought as an ordinary sigma model on $G/T$
represented by $\phit_{\m}$  and some extra abelian excitations,
$\phit_{\t}$ associated with $LT/T$. We have previously emphasized that
the abelian excitations do not contain a constant mode.
The gaussian integration of the above  quadratic action yields
\bea
	&&\exp
	\left[ -2\pi k\tau_2 \; {K(\cmu,\cmu) \over K(h_\psi,h_\psi)}
	\; + 2\pi\tau_2 \nu(\cmu)
	\right]	\nonumber\\
\label{quad-partition}
&\times&\left[\prod_{\alpha\succ 0}
\det\left(\d_{\zb} +{1\over 2}2\pi i\alpha(\cmu)\right)\right]^{-1}
	\;\left[\vphantom{{1\over 2} \prod_{\alpha\succ 0}}
	\widehat{\det}(\d_{\zb})\right]^{-l/2}\, ,
\eea
where $\widehat{\det}$ indicates the omission of the $x$ translationally
invariant modes and $l$ is the rank of the group $G$.

As was discussed   after Eq. (\ref{left2}), we also have to take
into account the prefactors coming form the ``spin'' part of the
transformation law arising from both the circle action and the $T$ action
on the wave functions.  We will see that these prefactors are crucial in
turning the above into an analytic function of $\tau$ as required by
(\ref{susy-part}).

\subsection{Group  Action Around the Weyl Points}
\label{group-action}

In the partition function  $Z(\theta, \tau_1, \tau_2)$ defined in
Eq.~(\ref{partition}), the  operators   $e^{i\theta}$ and $e^{2\pi i
\tau_1 P_1}$ act on the wave functions at the end of the paths or more
precisely on the sections of the matter line bundles. We saw in
Section~\ref{lagrangian} that this action induces both a spin and an
orbital spin part in our discussion of the line bundle $\L_{(\nu,0)}$.
As discussed at the beginning of Section~\ref{review} we actually need
to work with the line bundle $\L_{(\lambda,k)}$. By mimicking the
derivation at the end of Section~\ref{lagrangian}, we will compute the
action of $S^1\times T\times U(1)$ on a section of $\L{(\lambda,k)}$
and find both a ``spin'' and an orbital part.  The former will appear
as a prefactor in the computation of the path integral while the latter
also determines the proper boundary conditions to use in the evaluation
of the determinants.

The general form of the left action on the sections was given in
Eq.~(\ref{left2}). Using the same notations we see that we have to
determine the representation matrix $\rho(t)$ where $t$ is  the
solution of $\ell g\x= g\x (g^{-1}\x \ell g\x) = g\x t^{-1}$
with $\ell\in T$ and $g(x)\in LG$. The prefactor will
simply be given by the computation of $\rho(t)$
on the line bundle $\L_{(\lambda,k)}$. To perform  this calculation
we require the  Lie algebra relation
\beq
\left[ (X,r),(Y,s) \right] =([X,Y],\phi(X,Y))\;,
\eeq
where $(X,r)\in {L\g} \oplus {\Bbb{R}}$
(the infinitesimal version of the $\lgt$ multiplication law).
The algebra cocycle is explicitly given by
\beq
\label{cocycle}
\phi(X,Y)={i\over\pi}\int_0^1 dx {K(X(x),{d\over dx}Y(x))\over
K(h_\psi,h_\psi)}\;.
\eeq
{}From the quantum field theory viewpoint it is useful to reexpress the
above relations directly in terms of the currents $J_X$ which were defined
in (\ref{currents}). Note that we view $J_X$ as an operator acting on a
Hilbert space. In the same spirit we can often view the group element
$(g,u)$ as an operator of the form $ue^{J_X}$ in the case $k=1$. We then
recover the group law as a relation between operators:
\bea
ue^{J_X}\,ve^{J_Y}&=&uve^{J_X+J_Y+{1\over 2}[J_X,J_Y]+\cdots}\\
&=&uve^{J_X+J_Y+{1\over 2}(J_{[X,Y]}+\phi(X,Y))+\cdots}\\
&=&uve^{{1\over 2}\phi(X,Y)+\cdots} e^{J_X+J_Y+{1\over 2}J_{[X,Y]}
+\cdots}\; ,
\eea
where we have used the current algebra (\ref{k-m}).
In computing the prefactor, remember that the character index is
localized on the fixed points of $S^1\times T$ in $LG/T$,  so we need
only to
know the behavior of the sections only around the elements $\ewaff$ of the
affine Weyl group $\waff$.  In
a neighborhood of the fixed points we can parametrize an element of
$LG/T$ by $\widetilde{g}_{w,\varphi}(x)= w \exp\left(2\pi i \cmu
x\right)e^{\widetilde{\varphi}(x)}$.  It is just a matter of algebra to
find
\bea
e^{i\theta}e^{-2\pi i\tau_1 P_1}\widetilde{ g}_{w,\varphi}\x
&=&we^{2\pi i\cmu x}e^{\widetilde{\varphi}'(x-\tau_1)}
e^{i(\tw-2\pi\tau_1\cmu)}\nonumber\\
&\times &\exp{\left[-2i{K(\cmu,\tw)\over \K}\right]} \nonumber \\
&\times &\exp{\left[2\pi i\tau_1{K(\cmu,\cmu)\over\K}\right]}
e^{-2\pi i\tau_1 P_1} \;,
\eea
where we have defined $\theta_w=w^{-1}\theta w$ and $\displaystyle
\widetilde{\varphi}'=e^{-2\pi i\cmu x}e^{i\tw}\widetilde{\varphi}
e^{-i\tw} e^{2\pi i\cmu x}$.  From the above we deduce two results.
Firstly we see that the boundary conditions on the fluctuations around the
fixed points are given by
\beq
\label{bcxi}
\widetilde{\varphi}(x+\tau_1,\tau_2)=\widetilde{\varphi}'(x,0)\, .
\eeq
Secondly the prefactor (or spin part) for a bundle with weight
$(\lambda,k)$ is:
\beq
\label{prefac}
e^{i\lambda(\tw-2\pi\tau_1\cmu)}
\exp{\left[k\left(2 i{K(\cmu,\tw)\over \K}
-2\pi i\tau_1{K(\cmu,\cmu)\over \K}\right)\right]}\;.
\eeq
It is worth mentioning that this factor is independent of the scale of the
scalar product. The inclusion of the superpartner does not modify the
discussion above.

In more mathematical terms, the line bundle $\L_{(\lambda,k)}$ is
induced from the representation $(0,\lambda,k)$ on $S^1 \times T \times
U(1)$ for the principal bundle $S^1 \times \wt{LG}$ over $(S^1 \times
\wt{LG})\bigl/ (S^1\times T \times U(1)) = LG/T$. Left translation by
$S^1\times T \times U(1)$ has an induced action on $\L_{(\lambda,k)}$. At
a fixed point of the action of $S^1 \times T$ given by the affine Weyl
coset $\ewaff T = w e^{2\pi i \cmu x} T$, the induced action is
multiplication by a complex number of modulus one on the line
$\L_{(\lambda,k)}$ at the coset $\ewaff T$. That number in terms of
$\lambda$, $k$ and $\ewaff$ is given by formula (\ref{prefac}). In other
words, the spin part of the action is the lift of left translation to the
line bundle.

\subsection{Determinants}

The evaluation of the determinants will follow the discussion given in
\cite{akmw.cmp,akmw.irvine}. Because we are working on a torus one can
associate eigenvalues with each of the first order differential operators
which appear in (\ref{quad-action}). In any reasonable regularization
scheme one will have a term by term cancellation between the fermionic
modes and the ``anti-holomorphic'' bosonic modes. This is guaranteed by the
existence of the supersymmetry. Thus only the ``holomorphic'' bosonic
sector contributes in a non trivial way. We would like to regulate the
determinants in such a way that holomorphicity is preserved.

The determinant of $\d_{\zb}$ is easily evaluated with the result
\beq
	\widehat{\det} \d_{\zb} = \eta(\tau)^2\;,
\eeq
where the Dedekind $\eta$-function is defined by
\beq
	\eta(\tau) = q^{1/24}\; \prod_{n=1}^\infty \left(1-q^n\right)\;.
\eeq
Since our lagrangian describes particle motion on $LG/T$ we note that there
are no ``pointlike particle'' modes\footnote{A pointlike particle mode
would be the time evolution of a constant loop.} associated with $T$
ensuring the absence of any dependence on $\tau_2$ in the determinant
$\widehat{\det} \d_{\zb}$. Had the pointlike particle modes associated with
$T$ been present then we would have found $\det'\d_{\zb}= 2\tau_2
\eta(\tau)^2$ as the result of the gaussian integration\footnote{The prime
means the omission only of the zero eigenvalue mode.}. We should carefully
keep track of all such extraneous factors because equation
(\ref{susy-part}) tells us that the final answer must be an analytic
function of $\tau$.

We will now carefully define
\beq
\det\left(\d_{\zb} +{1\over 2}2\pi i\alpha(\cmu)\right)
\eeq
in  a way which preserves holomorphicity in $\tau$ and guarantees
the correct periodicity in $T$.
By using the boundary condition one
easily determines that the eigenvalues are given by
\beq
	{\pi\over \tau_2}\;\left( m + n\tau -\zeta \right)
\eeq
where $m$ and $n$ are integers, and
\beq
	\label{det-zeta}
		\zeta =  {\alpha(\tw) \over 2\pi}
		- \alpha(\cmu) \tau \;.
\eeq
It is useful to study the following formal ratio of determinants
\beq
{\det\left(\d_{\zb} +{1\over 2}2\pi i\alpha(\cmu)\right) \over
	\det'\left(\d_{\zb}\right)} =
{\displaystyle \prod 	{\pi\over \tau_2}\;\left( m + n\tau -\zeta \right)
	\over
\displaystyle \mathop{{\prod}'}	{\pi\over \tau_2}\;\left( m + n\tau \right) }
\eeq
where the prime again means to eliminate the $m=0$, $n=0$ mode. Note
that the right hand side is formally an odd function of $\zeta$. The
above ratio may be written as
\beq
\label{W-sig1}
	{-2\pi \zeta \over 2\tau_2} \; \mathop{{\prod}'}
	\left( 1 - { \zeta \over m + n\tau}\right) \;,
\eeq
which is only formal since the product is divergent. We proceed in
two different ways. Firstly we note that (\ref{W-sig1}) is
essentially the definition of the Weierstrass $\sigma$-function. If we
employ some unspecified cutoff then it may be rewritten
as:
\bea
\det\left(\d_{\zb} +{1\over 2}2\pi i\alpha(\cmu)\right)
&=&  - 2\pi \eta(\tau)^2 \sigma(\zeta;\tau) \nonumber\\
		\label{W-sig2}
	&\times& \exp\left[ -\mathop{{\sum}'} {\zeta \over m+n\tau}
	-{1\over 2} \mathop{{\sum}'}
	\left( {\zeta \over m+ n\tau}\right)^2 \right]\;.
\eea
The entire issue revolves on how we handle the divergent sums.
Expression~(\ref{W-sig2}) already tells us that the ambiguity in different
regularizations is a quadratic polynomial in $\zeta$.

Secondly, this may also be seen differently by noticing that formally
\beq
	{\d^3 \over \d \zeta^3} \log
	\det\left(\d_{\zb} +{1\over 2}2\pi i\alpha(\cmu)\right)
\eeq
is finite without need for regularization. We  can define the
regularized determinant by the differential equation
\beq
\label{der3-det}
	{\d^3 \over \d \zeta^3} \log
	\det\left(\d_{\zb} +{1\over 2}2\pi i\alpha(\cmu)\right)
	= - \sum {2\over (m + n\tau - \zeta)^3}\;.
\eeq
The right hand side is an elliptic function, the derivative of the
Weierstrass $\wp$ function,  because the sum is uniformly convergent.
Integrating~(\ref{der3-det}) leads to an expression
which will be ambiguous by a quadratic polynomial in $\zeta$. In
conclusion we have a holomorphic regularization scheme which leads to
a polynomial ambiguity as in all renormalizable quantum field
theories. For our purposes it is more convenient to express the answer
in terms of $\vartheta$-functions\footnote{We use the $\vartheta$
function conventions of Mumford \cite{mumford.tataI}. The function
$\vth(\zeta,\tau)$ is odd in $\zeta$. Definitions of the Weierstrass
functions and the constant $\eta_1$ may be found in
\cite{whittaker-watson}.}.  The following identity
\beq
	-2\pi \eta(\tau)^2 \sigma(\zeta;\tau) =
	{ \vth(\zeta;\tau) \over \eta(\tau) } \;
	e^{\eta_1  \zeta^2}
\eeq
allows us to express the determinant as
\beq
\det\left(\d_{\zb} +{1\over 2}2\pi i\alpha(\cmu)\right)
= { \vth(\zeta;\tau) \over \eta(\tau) }\; e^{{\cal P}(\zeta)}\;,
\eeq
where ${\cal P}(\zeta)$ is a quadratic polynomial in $\zeta$. It is easy to
resolve the ambiguity in the definition of the determinant. We note that as
we go around a cycle in the maximal torus $T$ the variable $\zeta$ shifts
by an integer, see (\ref{det-zeta}). For example, in $SU(2)$ the change is
$2$ as we go around the cycle in $T$. Since we are interested in studying
integral representations of $\lgt$, it follows that the determinant should
be periodic under $\zeta\to \zeta+r$, where the integer $r$ is the greatest
common factor of the shifts when all cycles of $T$ are considered. The
function $\vth(\zeta;\tau)$ has the aforementioned periodicity; hence the
polynomial must be of the form ${\cal P}(\zeta) = 2\pi i p \zeta/r + {\rm
constant}$, where $p$ is an integer. The integer $p$ is zero because it is
natural to require the determinant to be odd under $\zeta \to -\zeta$ as
was previously remarked. Thus we conclude that $p=0$. The value of the
constant may be fixed by requiring that the $n=0$ mode of the determinant
reproduces the corresponding one for a point particle (up to the standard
central charge correction). In summary, we can choose ${\cal P}(\zeta)$ to
be an appropriate constant and all properties we require will be satisfied.
Thus we summarize by saying that the value of the determinant in question
is
\beq
\label{det-theta}
\det\left(\d_{\zb} +{1\over 2}2\pi i\alpha(\cmu)\right)
	= { \displaystyle -i\vth\left({\alpha(\tw) \over 2\pi}
		- \alpha(\cmu) \tau ;\tau\right) \over \eta(\tau) }\;.
\eeq

The mathematical interpretation of (\ref{det-theta}) is as follows. Let
$J(\Sigma)$ be the Jacobian of the complex torus $\Sigma$.
Each point of $J(\Sigma)$ gives an elliptic operator
$\db_\chi$ where $\db: \Lambda^0(\Sigma) \to \Lambda^{(0,1)}(\Sigma)$
and $\db_\chi$ is $\db$ coupled to the flat line bundle determined by
the character $\chi$ of $\pi_1(\Sigma) \to \Bbb{C}\backslash\{0\}$. The
operators $\db_\chi$ have index zero so we get a holomorphic map $\phi$
from $J(\Sigma)$ to $\Fzero$, the space of Fredholm operators of
index zero\footnote{Strictly speaking, $\Fzero$ is the space of index
zero Fredholm operators from a Sobolev space $H^1(\Sigma)$ to
$L^2(\Sigma)$.}. But $\Fzero$ is a complex manifold with a natural
holomorphic line bundle, the determinant line bundle\footnote{The
determinant line bundle of an appropriate space $S$ will be denoted by
$\DET(S)$.} $\DET(\Fzero)$, and a
natural section $s$ \cite{quillen.det,freed.sandiego}.
Now $\DET(\db) = \phi^*(\DET(\Fzero))$ is a holomorphic line bundle over
$J(\Sigma)$ with holomorphic section $\phi^*(s)$. If we take the simply
connected covering of $J(\Sigma)$ with coordinates $(\zeta,\tau)$, then
$\phi^*(s)$ pulled up to the covering is a section of a trivial bundle;
it is the $\vartheta$-function
\beq
	{-i\vth\left(\zeta;\tau\right)\over \eta(\tau)}
\eeq
with transformation law (\ref{theta-X}) below. We are restricting the family of
$\db$ operators to those with $\zeta= \alpha(\tw)/2\pi
-\alpha(\cmu)\tau$. Hence we get the {\it function\/}
\beq
	{\displaystyle -i\vth\left({\alpha(\tw) \over 2\pi}
		- \alpha(\cmu) \tau ;\tau\right) \over \eta(\tau) }\;.
\eeq

\subsection{The Character Index  Formula}

We are now in a position to put the prefactors and the determinants
together in a concise expression for the character index
$I_{(\nu,k)}(\theta,\tau) = Z(\theta,\tau)$ of the Dirac-Ramond operator.
The reader is
reminded that one has to sum over $\waff$, the fixed points of the
$S^1 \times T$ action and that $\waff = W \ltimes \check{T}$. Collating
all the information developed in this section we have
\bea
	I_{(\nu,k)}(\theta,\tau) &=& \sum_{w \in W} \sum_{\cmu\in
	\check{T}} q^{\displaystyle k\,K(\cmu,\cmu)/K(h_\psi,h_\psi)}
	q^{\displaystyle -\nu(\cmu)}
	\nonumber\\
	&\times& \exp\left( i\nu(\tw) -2ik {K(\cmu,\tw) \over
	K(h_\psi,h_\psi)} \right) \nonumber\\
		\label{index0}
	&\times&	\left[\prod_{\alpha\succ 0}
\det\left(\d_{\zb} +{1\over 2}2\pi i\alpha(\cmu) \right)\right]^{-1}
	\;\left[\vphantom{{1\over 2} \prod_{\alpha\succ 0}}
	\widehat{\det}(\d_{\zb})\right]^{-l/2}\;.
\eea
Before expressing (\ref{index0}) in terms of classical functions we make
several remarks about the abstract structure. As we have noted, the
troublesome determinant we discussed is not a function but a holomorphic
section of a line bundle over the Jacobian variety of $\Sigma$. The
nontriviality of this section is closely related to the necessity for
regularization. We will presently see that the shift by the Coxeter number
arises because we have a section and not a function. Had the determinant
been finite (which of course it cannot) the shift by the Coxeter number
would not be there. Formula~(\ref{index0}) may be rewritten as,
\bea
	I_{(\nu,k)}(\theta,\tau) &=& \sum_{w \in W} \sum_{\cmu\in
	\check{T}} q^{\displaystyle k\,K(\cmu,\cmu)/K(h_\psi,h_\psi)}
	q^{\displaystyle -\nu(\cmu)}
	\nonumber\\
	&\times& \exp\left( i\nu(\tw) -2ik {K(\cmu,\tw) \over
	K(h_\psi,h_\psi)} \right) \nonumber\\
	\label{index1}
&\times&
	\left[\prod_{\alpha\succ 0}
	{ \displaystyle -i\vth\left({\alpha(\tw)\over 2\pi}
		- \alpha(\cmu) \tau ;\tau\right) \over \eta(\tau)
	}\right]^{-1} \; 	\eta(\tau)^{-l}\;,
\eea
and is the central result of this article. It is the natural
form for the character index from the path integral viewpoint. All
other forms are derived from this one by $\vartheta$-function
identities and algebraic manipulation.  There are several important
remarks which should be made before proceeding. As was strongly
advertised, (\ref{index1}) is an analytic function of $q$ which
involves $\vartheta$-functions on the Jacobian of $\Sigma$ and not
$\Theta$-functions on the torus $T$. As expected, the expression is
independent of the choice of scale for the inner product.
Formula~(\ref{index1}) is almost the Weyl-K\v{a}c
character formula; it is the index of the Dirac-Ramond
operator on $LG/T$ instead of the $\db$ operator on $LG/T$. As
explained in I, $\dbar$ and the Dirac operator are related by
twisting.  If we are interested in the character associated to a
representation with highest $G$ weight $\lambda$ then we should choose
$\lambda$ to differ from $\nu$ by the Weyl weight $\rho$.

To transform (\ref{index1}) into a more conventional form of the character
formula and to see the  shift by the Coxeter number we need
the $\vartheta$-function identity
\beq
	\label{theta-X}
	\vth(\zeta + m\tau;\tau) = (-1)^m q^{m^2/2} e^{-2\pi i m\zeta}
	\vth(\zeta; \tau)
\eeq
where $m$ is an integer and formula (\ref{coxeter}) for the Coxeter
number~$\cox$.  Thus the index may be rewritten as
\bea
 	I_{(\nu,k)}(\theta,\tau) &=&
	\sum_{w\in W} \sum_{\cmu \in \check{T}} q^{\displaystyle -\nu(\cmu)}
	q^{\displaystyle (k + \cox)K(\cmu,\cmu)/\K} \nonumber\\
	&\times& \exp\left( i\nu(\tw) -2i(k+\cox)\;
	{K(\cmu,\tw) \over \K} \right) \nonumber\\
\label{index2}
	&\times& \eta(\tau)^{-l}\;
	\left[\prod_{\alpha\succ 0}\;
	{ \displaystyle -i\vth\left({\alpha(\tw)\over 2\pi}
	;\tau\right) \over \eta(\tau)
	}\right]^{-1}\;.
\eea
We used the fact that, since $\rho$ is a weight
and $\cmu$ is in the coroot lattice, then $\rho(\cmu) \in {\Bbb Z}$.
Also, if we use the Killing form to identify $\t$ with $\t^*$, one
can easily see that $2\cmu/\K$ is a weight.

To write (\ref{index2}) in a more recognizable form, one proceeds in two
different ways. Either we do the $W$ sum first or we do the
$\check{T}$ one. These two alternatives lead to very different looking
formulas. We recall from I
that if $\nu$ is a weight of $G$ then the $T$-index of the Dirac
operator coupled to the $\nu$-line bundle is
\bea
	I_\nu(\theta) &=& \sum_{w\in W} e^{i\nu(\tw)}
	\prod_{\alpha\succ 0} {1\over 2i \sin {1\over 2} \alpha(\tw)}\\
	&=& \sum_{w\in W} (-1)^{\ell(w)} e^{i\nu(\tw)}
	\prod_{\alpha\succ 0} {1\over 2i \sin {1\over 2} \alpha(\theta)}\;,
\eea
where $\ell(w)$ is defined as
the number of positive roots turned into negative roots by $w$.
Note the ordinary index has only a single subscript while the
loop one has a double subscript.

It is convenient to define the Weyl denominator by
\beq
	D_W(\theta) = \prod_{\alpha\succ 0} 2i\sin\frac{1}{2}\alpha(\theta)
\eeq
and the K\v{a}c denominator by
\beq
	D_K(\theta) = \prod_{n>0} \left( 1- q^n\right)^l\;
	\prod_{\alpha\succ 0} \prod_{n>0}
	\left(1-q^n e^{i\alpha(\theta)}\right)
	\left(1-q^n e^{-i\alpha(\theta)}\right) \;.
\eeq
The Weyl and the K\v{a}c denominators are closely  related to our
index formulas because of the identity
\beq
	{-i \vth(\zeta;\tau) \over \eta(\tau)} = q^{1/12}\; 2i\sin\pi\zeta \;
	\prod_{n>0} \left(1 - q^n e^{2\pi i \zeta}\right)
		\left(1 - q^n e^{-2\pi i \zeta}\right) \;.
\eeq

It is now a matter of algebra to transform (\ref{index2}) into one of
the standard forms for the Weyl-K\v{ac} character formula.  Define the
sublattice  $\check{T}^* = \{ 2\cmu/\K\;|\;\cmu\in\check{T} \}$ of the
weight lattice, and the dilated-translated lattice $\check{T}^*(\nu,a) =
\nu+ a\check{T}^*$.  Now let us express (\ref{index2}) in a different
way by first summing over the Weyl group $W$. This organizes the
elements of the expansion in terms of the ordinary Dirac index:
\bea
I_{(\nu,k)}(\theta,\tau) &=& { q^{-(\dim\g)/24} \over D_K(\theta)}\;
	q^{\displaystyle - (\nu,\nu)/[(k+\cox)(\psi,\psi)]} \nonumber\\
	&\times& \sum_{\omega\in\check{T}^*(\nu,k+\cox)} I_{\omega}(\theta)
	q^{\displaystyle (\omega,\omega)/[(k+\cox)(\psi,\psi)]}\;.
\eea
Next, we could have first summed over the coroot lattice
generating a $\Theta$-function. Consider the lattice
$\check{T}^*(\nu,a)$
and the associated $\Theta$-function
\beq
\Theta_{(\nu,a)}(z,\tau) = \sum_{\omega\in \ct^*(\nu,a)}
	\exp\left[ {2\pi i\tau\over a}\;
	{\left(\omega,\omega\right) \over (\psi,\psi)}
	+ 2\pi i \omega(z) \right]\;.
\eeq
The index may be written as
\bea
	I_{(\nu,k)}(\theta,\tau) &=& {q^{-(\dim\g)/24} \over
	D_W(\theta) D_K(\theta)}\;
	q^{\displaystyle - (\nu,\nu)/[(k+\cox)(\psi,\psi)]}\nonumber\\
	&\times& \sum_{w\in W} (-1)^{\ell(w)}
	\Theta_{(\nu,k+\cox)}\left({\tw\over 2\pi},\tau\right)\;.
\eea

In order to incorporate the twist that turns the Dirac operator into $\dbar$
we remind you that in the Weyl character case
 the highest weight $\lambda$ is related to $\nu$
by $\nu = \lambda + \rho$ and that the character index and  group
character for $G$ are related by
\beq
	I_\nu(\theta) = \chi_{\subchi{\lambda}}(\theta)\;.
\eeq
Thus we see that the Weyl character formula for the highest weight
representation $\lambda$ of the group $G$ is
\bea
	\chi_{\subchi{\lambda}}(\theta) &=& \sum_{w\in W} e^{i(\lambda+\rho)(\tw)}
	\prod_{\alpha\succ 0} {1\over 2i \sin {1\over 2} \alpha(\tw)}\\
	&=& \sum_{w\in W} (-1)^{\ell(w)} e^{i(\lambda+\rho)(\tw)}
	\prod_{\alpha\succ 0} {1\over 2i \sin {1\over 2} \alpha(\theta)}\;.
\eea
In what follows we will write $\chi_{\subchi{\lambda}}$ even if $\lambda$ is
not a highest weight because every weight $\lambda$ is conjugate via
an element of the Weyl group to a highest weight.

In the same way, the loop index and the associated character for $\lgt$
are related by
\beq
	I_{(\nu,k)}(\theta,\tau) = \chi_{\subchi{(\lambda,k)}}(\theta,\tau)\;.
\eeq
A little algebra leads to the following two formulas for the character
\bea
	\chi_{\subchi{(\lambda,k)}}(\theta,\tau) &=&
	{ q^{-(\dim\g)/24} \over D_K(\theta)}\;
	q^{\displaystyle
	-[(\lambda+\rho,\lambda+\rho)-(\rho,\rho)]/[(k+\cox)(\psi,\psi)]}
	\nonumber\\
	\label{char1}
&\times& \sum_{\omega\in\check{T}^*(\lambda,k+\cox)}
	\chi_{\subchi{\omega}}(\theta)\;
	q^{\displaystyle [(\omega +\rho,\omega+\rho)-(\rho,\rho)]
	/[(k+\cox)(\psi,\psi)]} \;,\\
&=& {q^{-(\dim\g)/24} \over D_W(\theta) D_K(\theta)}\;
	q^{\displaystyle -
(\lambda+\rho,\lambda+\rho)/[(k+\cox)(\psi,\psi)]}
		\nonumber\\
	\label{char2}
	&\times& \sum_{w\in W} (-1)^{\ell(w)}
	\Theta_{(\lambda+\rho,k+\cox)}\left({\tw\over 2\pi},\tau\right)\;.
\eea
Equation (\ref{char1}) is the same as equation (14.3.10) of
\cite{pressley-segal.loopgroups} with the proviso
 that we use $L_0-c/24$
in our trace while they use $L_0$.
It is important to realize that in this context $c = \dim\g$,
see~(\ref{quad-action}), and that $c$ is not the Sugawara value.
Equation~(\ref{char2}) may be put in
a more useful form by mimicking the following computation with the Weyl
character formula. If one considers the trivial representation (highest
weight $\lambda=0$ with
$\chi_{\subchi{0}}(\theta)=1$) then one easily sees that the
Weyl denominator may be written as
\beq
	D_W(\theta) = \sum_{w\in W} (-1)^{\ell(w)} e^{i\rho(\tw)}
\eeq
and thus the Weyl character formula may be rewritten as
\beq
	\chi_{\subchi{\lambda}}(\theta) =
	{ \displaystyle \sum_{w\in W} (-1)^{\ell(w)} e^{i(\lambda+\rho)(\tw)}
	\over \displaystyle
	 \sum_{w\in W} (-1)^{\ell(w)} e^{i\rho(\tw)} }\;.
\eeq
The analogous equation in the loop group case exploits the fact that the
trivial representation has $\lambda=0$, $k=0$, and
$\displaystyle \chi_{\subchi{(0,0)}}(\theta,\tau) =q^{-(\dim\g)/24}$.
Thus we conclude that the denominators satisfy
\beq
	D_W(\theta) D_K(\theta) =
		q^{\displaystyle -(\rho,\rho)/[\cox(\psi,\psi)]}
	\sum_{w\in W} (-1)^{\ell(w)}
		\Theta_{(\rho,\cox)}\left({\tw\over 2\pi},\tau\right)\;,
\eeq
leading to the following form for the character formula
%\newpage
\bea
	\chi_{\subchi{(\lambda,k)}}(\theta,\tau) &=&
	q^{\displaystyle
	-(\lambda+\rho,\lambda+\rho)/[(k+\cox)(\psi,\psi)]}
	\nonumber\\
&\times& {\displaystyle \sum_{w\in W} (-1)^{\ell(w)}
	\Theta_{(\lambda+\rho,k+\cox)}\left({\tw\over 2\pi},\tau\right)
	\over \displaystyle \sum_{w\in W} (-1)^{\ell(w)}
		\Theta_{(\rho,\cox)}\left({\tw\over 2\pi},\tau\right)	}
\eea
which may explicitly obtained from the formulas in Chapter~12 of
\cite{kac.book} as discussed in the vicinity of equation~(A.25) in
reference~\cite{gepner-witten.wzw}.

It is well known that the affine characters have modular transformation
properties \cite{kac.book,pressley-segal.loopgroups}. The origin of these
properties was originally considered very mysterious but the connection of
affine Lie algebras to conformal field theory demystified the issue. In
\cite{gepner-witten.wzw}, the authors discussed the modular invariance of
the WZW model's partition function, a sum of the modulus squared of
characters. We can use our results to discuss the origin of the modular
properties of individual characters. The key observation is that the
quadratic action (\ref{quad-action}) is a non-chiral conformal field
theory. One should view the determinants in (\ref{index0}) as  short
hand for the path integral over the quadratic action.
The modular transformations properties of this chiral conformal field
theory explains the modular properties of the characters.

\par\vskip .2in\noindent
{\large\bf Acknowledgements}

We would like to thank C.~Itzykson for insisting that the ``proof of the
pudding is in the writing''. Each author would like to thank the home
institutions of the other two authors for visits while the research was
in progress.

\appendix
\section{Notational Conventions}

Let $G$ be a connected, simply connected,
compact simple Lie group with Lie algebra $\g$.
Let $T$ be its maximal torus with Lie algebra $\t$. The adjoint
actions by the element $x\in \g$ is denoted by $\ad x$ and is defined
by $(\ad x)y= [x,y]$ for $y\in\g$. The (negative definite) Killing
form is defined by
\beq \label{killing}
	K(x,y)= \Tr(\ad x \ad y)\;.
\eeq
The Killing form guarantees an orthogonal decomposition $\g = \t \oplus
{\euf m}$.
Here and in what follows a subscript $\t$ or $\m$ indicates the
projection along ${\euf t}$ or ${\euf m}$.
A root $\alpha$ is an element of the vector space $\t^*$
dual to
$\t$ which satisfies $(\ad t) e_\alpha = \alpha(t) e_\alpha$ where $t\in
\t$ and $e_\alpha$ are the ``raising'' and ``lowering'' generators. The set
of  roots will be denoted by $\Delta$. If the root $\alpha$ is positive
then we will write $\alpha \succ 0$. We are implicitly working in the
complexification of the Lie algebra. We use the Killing form to associate
elements of $\t$ with elements of $\t^*$. Our notation is as follows: given
$\beta\in \t^*$ one associates $t_\beta\in \t$ by the standard relation
\beq
	\beta(h) = K(h,t_\beta), \quad \forall h\in\t\;.
\eeq
The Lie algebra commutation relations may be written as
\bea
	[h, e_\alpha] &= &\alpha(h) e_\alpha \;,\\ \relax
	[e_\alpha, e_{-\alpha}] &= &K(e_\alpha, e_{-\alpha}) t_\alpha\;,
\eea
for $h\in\t$. It is convenient to find the ``standard'' ${\euf su}(2)$
subalgebras.  Define the inner product $(\cdot,\cdot)$ on
$\t^*$ by $ (\alpha,\beta) = K(t_\alpha,t_\beta)$. If one
defines {\it coroots} $h_\alpha$ by
\beq
	h_\alpha = {2 t_\alpha \over (\alpha,\alpha)}
\eeq
then
\bea
	[h_\alpha, e_{\pm\alpha}] &= \pm 2 e_{\pm\alpha}\;, \\ \relax
	[e_\alpha, e_{-\alpha}] &= h_\alpha\;.
\eea
The {\it coroot lattice\/}, $\check{T}$, is the integral lattice
spanned by the $h_\alpha$.  Elements of the coroot lattice will
usually be denoted by a ``check'' accent, \eg\ $\check{\mu}$. For
future reference notice that
\bea
	K(e_\alpha, e_{-\alpha}) &= &2/(\alpha,\alpha) \;,\\
	K(h_\alpha,h_\alpha) &= &4/(\alpha,\alpha) \;,\\
	\exp(2\pi i h_\alpha) &= &I \;.
\eea

Casimir operators and other quadratic objects will constantly appear in our
formulas and for this reason we offer a compendium of formulas. Let
$\{e_a\}$ be an arbitrary basis for $\g$ and let $K_{ab} = K(e_a,e_b)$ and
$K^{ab}$ be the inverse matrix then:
\begin{enumerate}
\item If $k_1,k_2 \in \t$ then
\beq	\label{cartan1}
	K(k_1,k_2)= \sum_{\alpha\in\Delta} \alpha(k_1)\alpha(k_2)\;,
\eeq
	see also (\ref{cartan2}).
\item The quadratic Casimir operator for an irreducible
	 representation $R$ of $\g$ is
	defined by
\beq
	C(R) = K^{ab}R(e_a) R(e_b)\;.
\eeq
\item The trace normalization $T(R)$ for an irreducible representation is
defined by
\beq
	\Tr R(e_a) R(e_b) = T(R) K_{ab}\;.
\eeq
	In our conventions we have $T(\ad)= 1$.
\item The following well known relation exists
\beq
	T(R) = { \dim R \over \dim \g}\;C(R)\;.
\eeq
It follows that $C(\ad) =1$.
\item Let $\{ \mu \}$ be the set of weights of a representation included
in the set according to multiplicity\footnote{If a weight has multiplicity
two then it appears twice in the set.} then
\beq
	C(R)= {\dim \g \over \dim R}\; {1\over \rank \g} \;
		\sum_{\{\mu\}}(\mu,\mu)\;.
\eeq
Applying this to the adjoint representation leads to the formula
\beq
\label{rank1}
	\rank \g= \sum_{\alpha\in\Delta} (\alpha,\alpha)\;.
\eeq
\item	If $\lambda$ is the highest weight of $R$ then
\beq
	{ C(R) \over(\lambda,\lambda)} \; =
	1 + { 2 \ip{\lambda}{\rho} \over \ip{\lambda}{\lambda} }
	\;  =
{(\lambda+\rho,\lambda+\rho)-(\rho,\rho) \over (\lambda,\lambda)}\;,
\eeq
where the {\it Weyl weight\/} $\rho$ is defined by $2\rho =
\sum_{\alpha>0} \alpha$.  Note that both the left hand side and the
right hand side of the above are independent of the scale of the inner
product. If one applies the above to the adjoint representation with
highest root $\psi$, which we will take to be a long root, then one
finds that the (dual) Coxeter number $\cox$ is given by
\bea
\displaystyle \cox \equiv {C(\ad) \over \ip{\psi}{\psi}} &=
	&\displaystyle 1 + {2 \ip{\psi}{\rho} \over \ip{\psi}{\psi}}\\
\label{coxeter}
\displaystyle = {1\over \ip{\psi}{\psi}} &=
	&\displaystyle K(h_\psi,h_\psi)/4\;.
\eea
Note that $2\ip{\psi}{\rho}/\ip{\psi}{\psi} = \rho(h_\psi)$ and thus
the  Coxeter number is an integer since $\rho$ is a weight. Also
notice that some of the above formulas are  independent of the scale
chosen for the inner product. Other formulas just express these
``scale invariant'' quantities in a specific choice of inner product.
\item The above observation leads to a scale invariant way of writing
(\ref{cartan1}). Note that the right hand side of (\ref{cartan1}) is
independent of the scale of the inner product while the left hand side
is not. We may rewrite (\ref{cartan1}) as
\beq
	\label{cartan2}
	4\cox\; { K(k_1,k_2) \over \K} = \sum_{\alpha\in\Delta}
		\alpha(k_1) \alpha(k_2)\;.
\eeq
\item Note also that (\ref{rank1}) may also be written in a scale
invariant way
\beq
	\rank \g= {1\over \cox}\;{\sum_{\alpha\in\Delta} (\alpha,\alpha)
		\over (\psi,\psi)}\;.
\eeq
\item Another useful formula is the ``strange formula'' of Freudenthal
which states that $(\rho,\rho) = (\dim\g)/24$ and may be written in a
scale invariant way as
\beq
	{ \dim\g\over 24} = {1\over \cox}\;{ (\rho,\rho) \over (\psi,\psi)}\;.
\eeq
\item The generator of $H^3(G, \Bbb{Z})$ is
\beq
\label{H3G}
\sigma =
{\ip{\psi}{\psi} \over 48 \pi^2} \; \Tr_{\rm ad} ( g^{-1}dg)^3=
{1 \over 12 \pi^2 K(h_\psi,h_\psi)} \; \Tr_{\rm ad} ( g^{-1}dg)^3\;,
\eeq
where the trace is taken in the adjoint representation of $\g$. The
last expression demonstrates that the generator is independent of the
normalization chosen for the inner product as expected.
\end{enumerate}

\small
%% \bibliographystyle{oapw}
%% \bibliography{alexandria}

\end{document}